\documentclass[prd,twocolumn,english,preprintnumbers,amsmath,amssymb,nofootinbib,a4]{revtex4-1}

\usepackage[latin1]{inputenc}
\usepackage{graphicx}
\usepackage{bbm}
\usepackage{amssymb}
\usepackage{amsmath}

\usepackage{dsfont}

\def\0#1#2{\frac{#1}{#2}}

\def\s0#1#2{\mbox{\small{$ \frac{#1}{#2} $}}}

\newcommand{\tr}{\mathrm{tr}}
\newcommand{\trf}{\mathrm{tr}_{\rm F}}

\newcommand{\I}{\mathrm{i}}
\newcommand{\be}{\begin{eqnarray}}
\newcommand{\ee}{\end{eqnarray}}

\newcommand{\nn}{\nonumber }
\newcommand{\fslash}{\hspace*{-0.2cm}\slash }

\newcommand{\Nf}{N_{\text{f}}}

\newcommand{\Nc}{N}

\newcommand{\bfe}{\langle A_0\rangle}

\newcommand{\Td}{T_{\rm d}}
\newcommand{\Tc}{T_{\chi}}
\newcommand{\luv}{\lambda_{\psi}^{\rm UV}}
\newcommand{\lfp}{\lambda_{\psi}^{\ast}}
\newcommand{\dr}{d({\rm R})}   
\newcommand{\dF}{d({\rm F})}   
\newcommand{\dA}{d({\rm A})}   
\newcommand{\dc}{d({\rm C})}

\newcommand{\ksb}{k_{\rm SB}}
\usepackage{babel}
\makeatother

\begin{document}

\title{On the Relation of the Deconfinement and the Chiral Phase Transition\\ 
in Gauge Theories with Fundamental and Adjoint Matter}

\author{Jens Braun} 
\affiliation{Institut f\"ur Kernphysik (Theoriezentrum), 
Technische Universit\"at Darmstadt, Schlossgartenstra\ss e 2, D-64289 Darmstadt, Germany}
\affiliation{Theoretisch-Physikalisches Institut,
  Friedrich-Schiller-Universit\"at Jena, Max-Wien-Platz 1, D-07743 Jena, Germany}
\author{Tina K. Herbst}
\affiliation{Institut f\"ur Physik, Karl-Franzens-Universit\"at, A-8010 Graz, Austria}

\begin{abstract}
We study the relation of confinement and chiral symmetry {breaking in gauge theories 
with non-trivial center}, such as~SU($N$) gauge theories. To this end, we deform these gauge theories by
introducing an additional control parameter into the theory and by varying the representation of the quark fields. 
We then consider a large-$d({\rm R})$ expansion of the effective action,
where~$d({\rm R})$ denotes the dimension of the representation~R of the quark
fields. We show how our large-$d({\rm R})$ expansion can be extended in a systematic fashion and discuss
the effects of $1/d({\rm R})$-corrections on the dynamics close to the finite-temperature phase boundary. 
Our analysis of the fixed-point structure of the theory suggests that 
the order, in which the chiral and the deconfinement phase transition occur, is dictated by the representation
of the quark fields and by the underlying gauge group. 
{In particular, we find that the phase diagram in the plane spanned by the temperature
and our additional control parameter exhibits an intriguing phase structure for quarks in the fundamental
representation. For SU($N$) gauge theories with adjoint quarks, on the other hand, the structure
of this phase diagram appears to be less rich, at least in leading order in the~$1/d({\rm R})$-expansion.}
\end{abstract}

\maketitle

\section{Introduction}
For quantum chromodynamics (QCD) with~$N\!=\!3$ colors and two (light) quark flavors it has
been found in various studies that the phase transitions associated 
with chiral symmetry restoration and deconfinement lie remarkably close to each other, 
see e.~g. Refs.~\cite{Meisinger:1995ih,Karsch:2000kv,Karsch:2003jg,Mocsy:2003qw,Fukushima:2003fw,Megias:2004hj,Ratti:2005jh,%
Cheng:2006qk,Aoki:2006br,Aoki:2006we,Schaefer:2007pw,Braun:2009gm,Aoki:2009sc,Herbst:2010rf,Pawlowski:2010ht,Herbst:2012ht}. 
From a phenomenological point of view, this observation has important consequences for our understanding
of the dynamics in heavy-ion collisions as well as of the generation of hadron masses in
the early universe. In fact, a comprehensive picture of the dynamics close to
the finite-temperature phase boundary of QCD is required for a reliable description
of data from heavy-ion collision experiments~\cite{BraunMunzinger:2003zd}.

While the underlying mechanisms associated with the confinement of quarks are not yet fully
understood, we have a profound understanding of chiral symmetry breaking in gauge theories.
In fact, it is already known from Nambu--Jona-Lasinio (NJL) models~\cite{Nambu:1961tp,Nambu:1961fr} that
chiral symmetry breaking is indicated by strong quark self-interactions. 
In particular, the four-quark interactions play a prominent role since
they are directly related to the chiral order parameter by means of a Hubbard-Stratonovich transformation. 
Contrary to NJL-type models, however, the quark self-inter\-actions are not free parameters of QCD but
generated dynamically and driven to criticality by the gauge degrees of freedom. 
This can be understood in simple terms
from a renormalization group (RG) analysis of the fixed-point structure of four-fermion 
interactions in gauge theories~\cite{Gies:2002hq,Gies:2005as,Braun:2005uj,Braun:2006jd,Braun:2008pi,Braun:2009ns,Braun:2010qs,Braun:2011pp}.
Such a fixed-point analysis also allows for a computation of 
the chiral phase transition temperature as
a function of the flavor number~$\Nf$, see Refs.~\cite{Braun:2005uj,Braun:2006jd}. 
{The results are indeed} in very good agreement with those from lattice
simulations~\cite{Cheng:2006qk,Aoki:2006br,Aoki:2006we,Aoki:2009sc}.
Also in agreement with lattice studies, 
the deconfinement and chiral phase transition temperature have
been found to almost coincide within
such a first-principles RG setup~\cite{Braun:2009gm,Pawlowski:2010ht}. 
In Ref.~\cite{Braun:2011fw} it has then been shown that the (almost) coincidence of the chiral and deconfinement 
phase transition temperature can also be understood by means of an analysis of the influence of gluodynamics
{on the fermionic fixed-point structure.}

In particular with respect to the phase boundary of QCD at finite temperature and quark
chemical potential, the interrelation of the chiral and the deconfinement phase transition is currently
under debate, see e.~g. Ref.~\cite{McLerran:2007qj}. A systematic deformation of QCD represents a valuable strategy to gain 
important insights into this question. From the response of the theory to such a deformation, we may then learn something about 
the underlying mechanisms at work. For example, {varying the number of
quark flavors, the number of colors, or the current quark masses
indicates that the nature} of the two phase transitions clearly depends on these
parameters of the theory, see e.~g.
Refs.~\cite{Brown:1990ev,Karsch:2003jg,Schaefer:2008hk,Panero:2009tv,
Braun:2010cy,Strodthoff:2011tz,Dumitru:2012}.
More recently, the interplay of the chiral and deconfinement phase transition has been analyzed by varying the 
boundary conditions of the quark fields in the temporal direction~\cite{Gattringer:2006ci}. Such a deformation of the theory yields
so-called dual observables which relate the spectrum of the Dirac operator to the order parameter for 
confinement, namely the dressed Polyakov loop~\cite{Synatschke:2007bz,Bilgici:2008qy,Kashiwa:2008bq,Sakai:2008py,Bilgici:2009tx,%
Braun:2009gm,Fischer:2009wc,Fischer:2009gk,Fischer:2010fx,Zhang:2010ui,Mukherjee:2010cp,Gatto:2010qs}. 

In this work, we consider a deformation of QCD different from the ones named above. To be specific, we deform QCD
by varying the representation of the quark fields. For example, one may
consider quark fields in the adjoint
or in the fundamental representation. From such a straightforward deformation it is then possible to gain
further insights into the mechanisms close to the finite-temperature phase boundary of QCD. In fact, 
basic properties of the theory can change when we change the representation of the quark fields: for example, quarks
in the adjoint representation do not break the underlying center symmetry of the gauge sector, whereas the
center symmetry is broken explicitly for quarks in the fundamental
representation. Since center-symmetry 
breaking is connected to the question of quark
confinement~\cite{Greensite:2003bk}, one may expect
that a variation of the representation of the quark field leaves its imprints in the phase structure of the theory.
In fact, it has been found in lattice simulations {of SU($N$) gauge theory
with adjoint quarks~\cite{Karsch:1998qj,Engels:2005te,Bilgici:2009jy} }
that the chiral phase transition temperature is significantly larger than the deconfinement phase 
{transition temperature, in contradistinction to QCD} 
with two (light) quark flavors in the fundamental representation. 
For a study of adjoint QCD with Polyakov-loop extended NJL (PNJL) models, we 
refer the reader to Refs.~\cite{Nishimura:2009me,Kahara:2012yr}. 

Here we pursue the strategy of Ref.~\cite{Braun:2011fw} and investigate
the fixed-point structure of quark self-interactions. On the one hand, the results of our study provide further
insights into the interrelation of quark confinement and chiral symmetry breaking. On the other hand, our fixed-point analysis
may help to develop new effective QCD low-energy models or to improve existing models. In this spirit, the present study also
extends previous works in which it has been discussed how the low-energy sector of QCD can be systematically connected
to the QCD Lagrangian at high momentum scales within a continuum
approach~\cite{Gies:2002hq,Gies:2005as,Braun:2005uj,Braun:2006jd,
Braun:2008pi,Braun:2009gm,Kondo:2010ts,Pawlowski:2010ht}.

This work is organized as follows: In Sect.~\ref{sec:GAintro} we discuss general aspects of our field-theoretical setup, with
an emphasis on the order parameters associated with quark confinement and chiral symmetry breaking. The fermionic
fixed-point structure is then discussed in detail in Sect.~\ref{sec:fermFP} for general representations of the quark fields.
In Sect.~\ref{sec:PBFP} we study the partially bosonized version of our fermionic ansatz discussed in Sect.~\ref{sec:fermFP}. 
The mapping between the two formulations is discussed in Sect.~\ref{sec:LNPBFP}. 
While the purely fermionic formulation of the matter sector already allows us to 
analyze the mechanisms at work at the phase boundary, the partially bosonized version 
allows us to gain access to the hadronic spectrum of the theory in a simple manner.
In Sect.~\ref{sec:BLNPBFP}, we then take into account $1/\dr$-corrections and discuss their effect on the
finite-temperature phase boundary. Our concluding remarks are given in Sect.~\ref{sec:conc}.

\medskip
\vfill

\section{General Aspects of QCD at Low Energies}\label{sec:GAintro}
In QCD phenomenology, the quarks are usually assumed to live in the fundamental representation
of the underlying~SU($N$) gauge group.
{In the construction of general gauge theories, however, the} 
quarks are by no means bound to live in the fundamental 
representation. In principle, they may transform according to any irreducible representation 
of the SU($N$) gauge group, e.~g. the fundamental representation ($N$-dimensional) or the adjoint 
 representation ($N^2\!-\!1$-dimensional). 
On the other hand, the gauge degrees of freedom always transform according to the 
adjoint representation of the gauge group.

In order to analyze the interplay of the chiral and the deconfinement phase transition in QCD, 
we deform QCD by varying the representation of the quark fields and by adding a relevant
coupling to the theory which can be considered as 
an external deformation parameter. For the latter, we choose a
four-fermion coupling~$\bar{\lambda}_{\psi}$: 
\be
S_{\rm QCD} \to S_{\rm QCD} + \int d^4 x\, \bar{\lambda}_{\psi} \left(\bar{\psi} {{\mathcal C}}\psi\right)^2\,,\nn
\ee
where $S_{\rm QCD}$ denotes the (classical) action of QCD, see e.~g.
Refs.~\cite{Kogut:1996mj,Catterall:2011ab} for lattice studies of this class of
theories. {The operator~$\mathcal C$ 
will be determined below. Loosely speaking, such a deformation allows us to ``detune" the chiral
and the confining dynamics of the theory. Due to the additional coupling,
the theory (``$\lambda_{\psi}$-deformed QCD") {now effectively depends on} two parameters, 
namely $\bar{\lambda}_{\psi}$ and $\Lambda_{\rm QCD}$.\footnote{Here, we assume
that the current quark masses are set to zero.} In particular, the values of chiral low-energy observables, 
depend on these two parameters. Different values of~$\bar{\lambda}_{\psi}$ can then be related to different
values of a given low-energy observable, such as the pion decay
constant~$f_{\pi}$. For~{$\bar\lambda_{\psi}\equiv 0$},
we are left with real QCD and the only input parameter is given by~$\Lambda_{\rm QCD}$ or, equivalently,
by the value of the strong coupling $\alpha_{\rm s}$ at some (high) momentum
scale. In this case,~$\Lambda_{\rm QCD}$ solely sets the scale for all physical 
observables~$\mathcal O$:~$\mathcal O\sim \Lambda_{\rm QCD}$.\footnote{One
could also turn the argument around and, for example, fix the scale by choosing
a certain value~$\Td$ for the deconfinement phase transition temperature. 
The latter then determines the scale~$\Lambda_{\rm QCD}$.}

In the following we would like to exploit the dependence on the two parameters~$\Lambda_{\rm QCD}$ and~$\bar{\lambda}_{\psi}$
to gain insights into the relation of the chiral and the deconfinement phase transition. To this end, we set up a 
model which shares many aspects with the full theory but can also be analyzed analytically to a large extent. Following 
Ref.~\cite{Braun:2011fw}, this allows us to come up with a prediction for a phase diagram 
in the plane spanned by the temperature and the pion decay constant~$f_{\pi}$ for
different representations of the quark fields. This prediction can then be tested with the aid of other approaches, such as
lattice simulations. 

{We would like to mention that} gauge theories with an additional relevant parameter, such as a four-fermion coupling,
have also attracted a lot of attention in recent years in beyond standard model applications, see e.~g. 
Refs.~\cite{Fukano:2010yv,Catterall:2011ab}. In particular, the case of SU($2$) gauge theory with two adjoint
quarks is of interest, see e.~g. Refs.~\cite{DelDebbio:2008zf,Catterall:2008qk,DelDebbio:2009fd,Sannino:2009za}.
{With regard to our present study, a word of caution needs to be added at this point. 
In our numerical analysis in Sects.~\ref{sec:fermFP} and~\ref{sec:PBFP}
we mostly restrict ourselves to the case of SU($2$) gauge theory with two massless adjoint quarks. We are aware of the
fact that this theory could already lie in the conformal window. In this work, however, we assume that 
the zero-temperature ground state of SU($2$) gauge theory with two adjoint quarks is governed by dynamical 
chiral symmetry breaking which would indeed appear to be the case within our present 
approximations. This is in accordance with Refs.~\cite{Dietrich:2006cm,Fukano:2010yv}.
As a first step, it is therefore natural for us to consider this theory in our numerical studies.
Even if our present approximations should turn out to be insufficient to describe correctly the chiral 
ground-state properties of SU($2$) gauge theory with two massless adjoint quarks, we still expect that our results will be similar 
for gauge groups of higher rank and broken chiral symmetry in the zero-temperature limit, see 
also our discussion in Sects.~\ref{sec:fermFP} and~\ref{sec:PBFP}.

{Before we now study} the interplay of the chiral and the deconfinement phase
transition in detail, we summarize a few field-theoretical aspects and
explain the general setup which underlies our study.
\subsection{Gauge Sector}\label{sec:GA}
Since we are interested in an analysis of the relation of the deconfinement and
chiral phase transition in general gauge theories (with non-trivial center), such as 
SU($N$) and Sp($N$) gauge theories, we need to discuss at least
some properties of the order parameters for confinement and chiral symmetry breaking.
In the following we first present a few important analytic relations
for the confinement order parameter. These relations will play an important role in our 
classification of gauge theories in the remainder of this work. 

The deconfinement phase transition in pure SU($N$) gauge theories has been studied in great detail.
For example, results are available from lattice simulations, see e.~g.
Refs.~\cite{Cheng:2006qk,Aoki:2006br,Aoki:2006we,Aoki:2009sc,Panero:2009tv,Cheng:2009zi,Datta:2010sq,%
Borsanyi:2010zi,Bazavov:2010pg,Kanaya:2010qd,Bornyakov:2011yb,Maas:2011ez}, as well as from functional
continuum methods~\cite{Braun:2007bx,Marhauser:2008fz,Braun:2010cy}. A well-known order parameter
for the deconfinement phase transition is the Polyakov loop. The associated Polyakov-loop
variable reads
\be
L_{\rm F}[A_0]=\frac{1}{\dF}\, {\mathcal P}\,{\rm e}^{ {\rm i}\bar{g}\int_0^{\beta} dx_0\,A_0(x_0,\vec{x}) }
\,,
\ee
where $\beta=1/T$ is the inverse temperature, $\dF$ is the dimension of 
the fundamental representation of the gauge group (e.~g. $\dF=\Nc$ for SU($N$)
gauge theories), $\bar{g}$~denotes the bare gauge coupling and $\mathcal P$
stands for path ordering.
The Polyakov loop is then given by~$\langle \tr_{\rm F} L_{\rm F} \rangle$.

Strictly speaking, the Polyakov loop $\langle \tr_{\rm F} L_{\rm F}\rangle$ is an order parameter
for center symmetry breaking, see e.~g. Ref.~\cite{Greensite:2003bk}.
However, its logarithm can also be viewed as half of the free energy~$F_{q\bar{q}}$ of a quark-antiquark pair 
at infinite distance.
A center-symmetric confining phase is signaled by a vanishing Polya\-kov
loop and implies that the free energy of a static fundamental quark (fundamental color source)~$F_q\simeq (1/2)F_{q\bar{q}}$
is infinite. The associated quark-antiquark potential is linearly rising for large
distances in this phase and no string breaking occurs. 
On the other hand, the deconfined phase is associated with a finite free energy~$F_q$ and
a finite Polyakov loop, i.~e. (spontaneously) broken center symmetry, see below.

Apart from the Polyakov loop, other order parameters for quark confinement have been
introduced, such as dual observables~\cite{Gattringer:2006ci}. In this work, however, we shall mainly consider
an order parameter which is closely related to the standard Polyakov loop, 
namely~${\rm tr_F}L_{\rm F}[\langle A_0\rangle]$. In Polyakov-Landau-DeWitt gauge it has indeed been shown
that the quantity~${\rm tr_F}L_{\rm F}[\langle A_0\rangle]$ serves as an order parameter for confinement~\cite{Braun:2007bx,Marhauser:2008fz}.
Here, $\bfe$ is a constant element of the Cartan subalgebra of the gauge group and denotes the 
ground state of the order parameter potential in the adjoint algebra, namely the so-called
Polyakov-loop potential.\footnote{Strictly speaking, we have to distinguish
between the background temporal gauge field and its expectation value~$\bfe$
associated with the order parameter for confinement,~${\rm tr_F}L_{\rm
F}[\langle A_0\rangle]$. We skip this subtlety here and refer to $\bfe$ as the
position of the ground-state of the order-parameter potential if
not indicated otherwise.} 
In a one-loop approximation this order parameter potential has first
 been computed in Refs.~\cite{Weiss:1980rj,Gross:1980br}. Based on a functional RG approach, a 
non-perturbative study of this potential, including a computation of the phase 
transition temperatures for several gauge groups, has first been carried out in
Refs.~\cite{Braun:2007bx,Marhauser:2008fz,Braun:2010cy}.
{Since the phase transition temperature of a given gauge theory represents a physical
quantity, it can be easily compared to results from other approaches, such as lattice
gauge theory, see e.~g. Refs.~\cite{Panero:2009tv,Datta:2010sq,Bazavov:2010pg,Kanaya:2010qd}.
For example, such a comparison shows that}
the RG result for the deconfinement phase transition temperature~$\Td$ for SU($3$) is in very
good agreement with results from lattice simulations.\footnote{{The deconfinement phase transition has also been
studied with matrix models, see e.~g. Refs.~\cite{Pisarski:2000eq,Dumitru:2003hp,Dumitru:2010mj,Dumitru:2012}. 
Based on input from lattice simulations, on the other hand, ways to improve} Polyakov-loop potentials widely used in PNJL/PQM-type model 
studies have been discussed in a recent review, see Ref.~\cite{Fukushima:2011jc}. For recent progress in this direction, we 
refer to Refs.~\cite{Sasaki:2012bi,Ruggieri:2012ny}. We emphasize that our present study relies directly on the order-parameter 
potential spanned by the background temporal gauge field. The position~$\bfe$ of the ground state of this potential appears in the 
Feynman diagrams associated with the dynamics in the matter sector, see discussion  in
Sects.~\ref{sec:setupmatter},~\ref{sec:fermFP} and~\ref{sec:PBFP}.}

In the following we shall refer to the temperature~$\Td$ as the
deconfinement phase transition temperature, even if we study theories with quarks in a 
representation other than the fundamental representation. As we shall briefly discuss below,
there exist representations for which the associated quark-antiquark potential
is not linearly rising at large distances, independent of the temperature~$T$.
Below~$\Td$, however, 
center symmetry is restored also in these cases. In any case, free (static) color charges
are always screened to form color-neutral states below~$\Td$ for any representation that is
considered in this work.

Let us now discuss some important properties of the quantity 
\be
L_{\rm R}[A_0]=\frac{1}{\dr}\, {\mathcal P}\,{\rm e}^{ {\rm i}\bar{g}\int_0^{\beta} dx_0\,A_0(x_0,\vec{x}) }\,.
\ee
Here, R~denotes the representation of the matter fields, e.~g. fundamental (F) or adjoint (A), and~$\dr$ is 
the dimension of the representation~R.  

{Under an arbitrary center transformation of the ground state~$\bfe$,}
\be
\bfe \to \bfe_{z}\,,
\ee
the quantity~${\rm tr_R}L_{\rm R}[\langle A_0\rangle]$ transforms as
\be
{\rm tr_R}L_{\rm R}[\langle A_0\rangle] \to z^{{\mathcal N}_{\rm R}} {\rm tr_R}L_{\rm R}[\langle A_0\rangle]\,,
\label{eq:poltrafoR}
\ee
where~${\mathcal N}_{\rm R}$ denotes the so-called $N$-ality of the representation~R.
For~SU($N$) gauge theories, for example, we have
\be
z\in\left\{ {\rm e}^{\frac{2\pi{\rm i}n}{N}}\right\}_{n=0,\dots\,,N-1}\,.
\ee
Moreover, we have~${\mathcal N}_{\rm R}=1$ for~R=F. We then obtain
\be
{\rm tr_F}L_{\rm F}[\langle A_0\rangle]\to z\, {\rm tr_F}L_{\rm F}[\langle A_0\rangle]
\ee
and conclude that~${\rm tr_F}L_{\rm F}[\langle A_0\rangle]$ is an order parameter for center symmetry
breaking and, loosely speaking, signals confinement of (static) quarks in the fundamental 
representation, see our discussion above. Note that~$\langle {\rm tr_F}L_{\rm F}[A_0] \rangle$ transforms accordingly
under arbitrary center transformations of the gauge field~$A_0$.
Hence, both ${\rm tr_F}L_{\rm F}[\langle A_0\rangle]$ and $\langle {\rm tr_F}L_{\rm F}[A_0] \rangle$ represent 
order parameters for center symmetry breaking. 

By construction,~$\bfe$ is an element of the Cartan subalgebra. 
In the following we may therefore parameterize it in terms of the generators of this
subalgebra:
\be
\beta \bar{g} \bfe &=& 2\pi\sum_{a=1}^{\dc}T^{(a)}\phi^{(a)}
=2\pi\sum_{a=1}^{\dc}T^{(a)}v^{(a)}|\phi|\,, 
\label{eq:hermmat}
\ee
where $v^2=1$, the $T^{(a)}$'s are the generators of the underlying gauge group
{in a given representation~R.} We shall refer to the set~$\{\phi^{(a)}\}$ as the coordinates
of~$\bfe$. 

The center-symmetric phase is signaled by~\cite{Marhauser:2008fz}
\be
{\rm tr_F}L_{\rm F}[\langle A_0\rangle] = \langle {\rm tr_F}L_{\rm F}[A_0] \rangle=0\,.\nn
\ee
In the class of Polya\-kov-DeWitt gauges, it then follows that the
position~$\bfe$ of the 
center symmetric ground state is uniquely determined by~\cite{Braun:2007bx,Braun:2010cy,Braun:2011pp}
\be
{\rm tr_F}(L_{\rm F}[\langle A_0\rangle]^{n})=0\,,\label{eq:polnfund}
\ee
where $n=1,\dots,\dc$ is the dimension of the associated Cartan subalgebra.
For SU($N$), we have~$\dc=N-1$.
At (asymptotically) high temperatures, on the other hand, we are in the perturbative regime 
where~$\bfe \to 0$ and~${\rm tr_F}L_{\rm F}[\langle A_0\rangle]\to 1$, 
see e.~g. Refs.~\cite{Weiss:1980rj,Gross:1980br,Braun:2007bx,Marhauser:2008fz,Braun:2010cy}.
Since~${\rm tr_F}L_{\rm F}[\cdot]$ is a monotonic function in the domain defined by 
the trajectory of~$\bfe$ as a function of the temperature~$T$, the quantity~${\rm tr_F}L_{\rm F}[\bfe]$ is 
monotonic and~${\rm tr_F}L_{\rm F}[\bfe]>0$ in the phase with broken center symmetry, see also our
discussion below.

The coordinates~$\{\phi^{(a)}\}$ of the {center-symmetric} ground state,
that are determined by Eq.~\eqref{eq:polnfund}, are
{given by ~$\{1/2\}$ for SU($2$) Yang-Mills theory and~$\{2/3,0\}$ for~SU($3$), respectively.}
From our discussion it is also clear that the order-parameter potential in the adjoint algebra is 
periodic.\footnote{The order-parameter potential is also invariant under
discrete rotations about the origin. The corresponding rotation angles {are determined by} the gauge group
under consideration.}
The lengths
of the periods in the various directions depend on the eigenvalues of the associated generators~$T^{(a)}$. 
Center transformations of the ground state~$\bfe$ can now be viewed as discrete rotations of 
the coordinates~$\{\phi^{(a)}\}$ around the center symmetric point. For example, we  
have a reflection symmetry with respect to~$\phi=1/2$ for SU($2$). The associated 
center transformation can then be written as follows:
\be
\phi \to \phi_z = 1 - \phi\,\label{eq:su2center}
\ee
with~$\phi \in [0,1/2]$.
Under such a center transformation, the order parameter transforms according to
\be
&& 
{\rm tr_F}L_{\rm F}[\bfe] \equiv {\rm tr_F}L_{\rm F}[\phi]= \cos\left( \pi \phi \right) \nn\\
&& \qquad\qquad\qquad 
\to {\rm tr_F}L_{\rm F}[\bfe_z] = - {\rm tr_F}L_{\rm F}[\bfe]\,,
\ee
as expected from Eq.~\eqref{eq:poltrafoR}. Note that we have used Eq.~\eqref{eq:hermmat} to 
express~$\bfe$ in terms of~$\phi$.

For SU($3$), center transformations of the ground-state~$\bfe$ can be written as rotations 
by angles of $2\pi n/3$ around the center-symmetric point~$\{2/3,0\}$, where $n=0,1,2$. Under such transformations, 
one then finds that the order parameter transforms as given in Eq.~\eqref{eq:poltrafoR}.

We would like to add that Eq.~\eqref{eq:polnfund} holds only for the center symmetric ground state~$\bfe$ 
for~$n \mod N \neq 0$. For $n \mod N=0$ and~$N$ even, we have
\be
{\rm tr_F}(L_{\rm F}[\langle
A_0\rangle]^{n})=(-1)^{\frac{n}{N}}\frac{1}{N^{n-1}}\,.
\ee
For odd~$N$ and~$n \mod N=0$, on the other hand, we have
\be
{\rm tr_F}(L_{\rm F}[\langle A_0\rangle]^{n})=\frac{1}{N^{n-1}}\,.
\ee

Let us now turn to representations other than the fundamental one.
From Eq.~\eqref{eq:poltrafoR}, we also observe 
that there may exist representations~R of the gauge group 
for which~${\rm tr_R}L_{\rm R}[\langle A_0\rangle]$
does {\it not} represent an order parameter for center symmetry breaking. To be specific, we 
consider~${\rm tr_R}L_{\rm R}[\langle A_0\rangle]$ for SU($N$) and~R=A (adjoint representation). 
Since we have~${\mathcal N}_{\rm R}=0$ (zero $N$-ality) in this case, {we find that
${\rm tr_{\rm A}}L_{\rm A}[\langle A_0\rangle]$ transforms as 
\be
{\rm tr_A}L_{\rm A}[\langle A_0\rangle]\to  {\rm tr_A}L_{\rm A}[\langle A_0\rangle]\,
\ee
under a center transformation.}
To be more specific, for SU($2$), we have
\be
\tr_{\rm A} L_{\rm A}[\bfe]\equiv \tr_{\rm A}L_{\rm A}[\phi] = \frac{1}{3}\left[ 1 + 2\cos\left(2\pi\phi\right)\right],
\ee
which is insensitive to arbitrary (center) transformations of~$\bfe$, see Eq.~\eqref{eq:su2center}.
Thus,~${\rm tr_A}L_{\rm A}[\langle A_0\rangle]$ is not an order parameter 
for center symmetry breaking, see also Refs.~\cite{Hubner:2007qh,Gupta:2007ax}. 

The insensitivity of~$\tr _{\rm A} L_{\rm A}[\bfe]$ and~$\langle \tr_{\rm A}L_{\rm A}[A_0]\rangle$ 
with respect to center transformations is related
to the fact that quarks in the adjoint representation do not break the 
underlying center symmetry of the gauge group, in contrast to 
quarks in the fundamental representation. From a phenomenological
point of view, there is indeed no strict notion of confinement of quarks in the
adjoint representation, even in the static limit. In this case, (static) quarks can be 
screened by the gluonic degrees of freedom and form a color-singlet state, as can be seen
from the decomposition of the tensor product of two adjoint multiplets.
Therefore a quark-antiquark pair at large distances can split up into two
singlet states. The associated quark-antiquark potential thus flattens at
large distances and does not rise linearly, as it is the case for (static)
quarks in the fundamental representation, see e.~g. 
Refs.~\cite{Karsch:1998qj,Engels:2005te,Hubner:2007qh,Gupta:2007ax,Bilgici:2009jy} 
for lattice studies. In particular, the 
Polyakov-loop~$\langle \tr_{\rm A} L_{\rm A}[A_0]\rangle$ is finite for all 
temperatures~\cite{Hubner:2007qh,Gupta:2007ax}. 
Since~$\langle \tr_{\rm A} L_{\rm A}[A_0]\rangle$ is related to the free energy of a static
(adjoint) quark, it follows that the free energy is finite, even in the center symmetric
phase at low temperatures. 
Note that the behavior of the quantities~$\langle \tr_{\rm A} L_{\rm
A}[A_0]\rangle$ and~$\tr_{\rm A} L_{\rm A}[\bfe]$ changes qualitatively
at~$T=\Td$, even though they do not represent order parameters for center
symmetry breaking. As we shall discuss below, this is due to the fact
that~$\langle\tr_{\rm A} L_{\rm A}[A_0]\rangle$ and~$\tr_{\rm A} L_{\rm
A}[\bfe]$ can be related to the order parameters~$\langle \tr_{\rm F} L_{\rm
F}[A_0]\rangle$ and~$\tr_{\rm F} L_{\rm F}[\bfe]$, respectively.

Let us close this subsection by summarizing a few useful relations for the 
quantity~$\tr_{\rm R} L_{\rm R}[\bfe]$. First, we note that the order 
parameter~${\rm tr_F}L_{\rm F}[\langle A_0\rangle]$ can be related to
the standard Polyakov-loop via the {\it Jensen inequality}. For a given
{\it concave} function~$f(\cdot)$, we have~$f(\langle\cdot\rangle)\geq \langle f(\cdot)\rangle$.
For example, this yields
\be
{\rm tr_F}L_{\rm F}[\langle A_0\rangle] \geq \langle {\rm tr_F}L_{\rm F}[A_0] \rangle\,
\label{eq:Jineq}
\ee
for~SU($2$) and SU($3$) gauge theory in the deconfined 
phase~\cite{Braun:2007bx,Marhauser:2008fz,Braun:2010cy}. We emphasize that this inequality
does not hold for general gauge groups and representations~R, since
it requires that ${\rm tr_R}L_{\rm R}[\cdot]$ is a concave function in the relevant domain.
Provided that~$\bfe(T)$ lies sufficiently close to the 
origin (e.~g. for sufficiently large temperatures~$T$), however, the inequality may hold for any gauge group
and representation.\footnote{In general, a subdomain around the origin can be found such that
${\rm tr_R}L_{\rm R}[\cdot]$ is a concave function. For a more detailed discussion of the
relation~\eqref{eq:Jineq} between the order parameters~${\rm tr_F}L_{\rm F}[\langle A_0\rangle]$
and~$\langle {\rm tr_F}L_{\rm F}[A_0] \rangle$, we refer the reader to Ref.~\cite{Braun:2011fw}.}

In addition to Eq.~\eqref{eq:Jineq}, 
we have the following two simple but useful inequalities:
\be
0\leq \frac{1}{\dr}\left|\tr_{\rm R}(L_{\rm R}[\bfe]^n)\right|\leq
\frac{1}{(\dr)^n}\,,
\ee
which follows from the generalized triangle inequality, and
\be
\!\!\!\!\!
-\frac{1}{(\dr)^n}\leq 
\frac{1}{\dr}\mathfrak{Re}\left[{\tr_{\rm R}(L_{\rm R}[\bfe]^n)}\right]\leq
\frac{1}{(\dr)^n}
\ee
for~$n\in\mathbb{N}$.

Finally, we evaluate the quantity~${\rm tr_R}L_{\rm R}[\langle A_0\rangle]$ for specific
configurations of the ground-state~$\bfe$. At very high temperatures~\mbox{$T\gg \Td$}, we have~$\bfe\to 0$ 
and~${\rm tr_R}L_{\rm R}[\langle A_0\rangle]\to 1$, independent of the gauge group and
the representation~R. In the low-temperature phase ($T<\Td$), however, the value
of~${\rm tr_R}L_{\rm R}[\langle A_0\rangle]$ depends on the gauge group and the 
representation~R. For example, we have~${\rm tr_F}L_{\rm F}[\langle A_0\rangle]=0$
for~SU($N$) gauge theories for~$T<\Td$. For the adjoint representation, on the other hand, 
we find
\be
\tr_{\rm A}(L_{\rm A}[\langle
A_0\rangle]^n)=-\frac{1}{(\dA)^n}=-\frac{1}{(N^2-1)^n}\label{eq:traconf}
\ee
with $n \mod N\neq 0$. For $n\mod N=0$, we have
\be
\tr_{\rm A}(L_{\rm A}[\bfe]^n) = \frac{1}{(d({\rm A}))^{n-1}}\,.
\ee
In SU($N$) gauge theories, the relation~\eqref{eq:traconf} follows straightforwardly 
from the fact that the tensor product of the triplet and the anti-triplet
can be decomposed into the adjoint multiplet and a singlet, $N\otimes\overline{N}=(N^2\!-\!1)\oplus 1$.
Using that the character of the product representation is given by the product of the characters
of the representations, we find
\be
\dA\,\tr_{\rm A}(L_{\rm A}[\bfe])
= \left|\dF\, \tr_{\rm F}(L_{\rm F}[\bfe])\right|^2 \!-\!1\,,
\ee
and similar relations for~$\tr_{\rm A}(L_{\rm A}[\bfe]^n)$ 
with~$n>1$ ($n\in\mathbb{N}$).

Up to this point, we have discussed that below~$\Td$ the quantity~$\tr_{\rm R}L_{\rm R}[\bfe]$
is zero for~R=F but negative for~R=A. Depending on the representation~R, however, it can also assume positive values in the
center symmetric phase. For example, let us 
{consider the ten-dimensional} representation (${\rm R}={\bf 10}$) of SU($3$). We then
find\footnote{Here, we have used that the tensor product of the triplet and the sextet can be 
decomposed into a singlet and decuplet. Moreover, the tensor product of two triplets can be
decomposed into a sextet and an anti-triplet.} 
\be
&&
d({\bf 10})\,\tr_{\rm {\bf 10}}(L_{\rm {\bf 10}}[ \bfe])
\nn\\&&\qquad\quad
= \left[\dF\, \tr_{\rm F}(L_{\rm F}[ \bfe])\right]^3  \nn\\
&& \qquad\qquad\quad
-2\left|\dF\, \tr_{\rm F}(L_{\rm F}[\bfe])\right|^2
+1 =1
\label{eq:polten}
\ee
for $T<\Td$, and similar relations for~$\tr_{\rm 10}(L_{\rm 10}[\bfe]^n)$ 
with~$n>1$ ($n\in\mathbb{N}$). Note that the
$N$-ality of the ten-dimensional representation of SU($3$) is zero as well,~${\mathcal N}_{\bf 10}=0$.
We add that also higher-dimensional representations exists for which we 
have~$\tr_{\rm R} L_{\rm R}[\bfe]=0$ for~$T<\Td$, even if the $N$-ality
of the representation is zero.

Depending on the representation~R, we have seen that the 
quantity~$\tr_{\rm R} L_{\rm R}[\bfe]$ can be zero, positive or negative in the center symmetric phase,
i.~e. for~$T<\Td$. This allows us to classify gauge theories. 
In Sect.~\ref{sec:fermFP}, we shall see that this classification
is to some extent related to the question whether the chiral phase transition temperature is larger or smaller 
than the deconfinement phase transition in a given gauge theory.

\subsection{Matter Sector}\label{sec:setupmatter}
Let us now discuss our field-theoretical setup in the matter sector.
Up to this point, our statements concerning the pure gauge sector are exact and can be obtained 
analytically. They only rely on the basic assumption that we work in the class
of Polyakov-DeWitt gauges. For our analysis of the matter sector, we
employ the following ansatz for the quantum effective action~$\Gamma$:
\be
\Gamma [\bar{\psi},\psi,\langle A_0\rangle]&=&\int d^4 x \Big\{ Z_{\psi}\bar{\psi}\left(
{\rm i}\partial\fslash + \bar{g}\gamma_0 \langle A_0\rangle\right)
\psi \nn\\
&& \quad + \frac{\bar{\lambda}_{\psi}}{2}\left[ (\bar{\psi}\psi)^2 
- (\bar{\psi}\vec{\tau} \gamma_5\psi)^2\right]\Big\}\,,
\label{eq:fermionic_action}
\ee
where~$Z_{\psi}$ is the wave-function renormalization of the quark fields.
In the present work, we restrict ourselves to $\Nf=2$ massless quark flavors with $d({\rm R})$
colors. The $\tau_i$'s represent the Pauli matrices and couple the spinors in flavor space. 

Our ansatz~\eqref{eq:fermionic_action} for the matter sector is {\it perturbatively}
non-renormalizable, as it is the case for the NJL model. Therefore we define it with
an UV cutoff~$\Lambda$ which then represents an additional parameter of the model. 
This setup also implies that the regularization scheme belongs to the definition of 
the model. The role of~$\Lambda$ for the fixed-point structure will be discussed
in detail in Sect.~\ref{sec:PBFP}.

For our study of the RG flow of the four-fermion coupling~$\bar{\lambda}_{\psi}$,
we employ the so-called Wetterich equation~\cite{Wetterich:1992yh}. The latter is 
an RG equation for the quantum effective action. In this approach, the effective action~$\Gamma$ 
depends on the RG scale $k$ (infrared cutoff scale) which determines the RG `time' 
$t=\ln(k/\Lambda)$ with $\Lambda$ being a UV cutoff scale. For reviews on
and introductions to this functional RG approach, we refer the reader to
Refs.~\cite{Litim:1998nf,Bagnuls:2000ae,Berges:2000ew,Polonyi:2001se,Delamotte:2003dw,%
Pawlowski:2005xe,Gies:2006wv,Schaefer:2006sr,Delamotte:2007pf,Rosten:2010vm,Braun:2011pp}.

Concerning the background field~$\langle A_0\rangle$, 
we will \emph{not} make use of the approximation~$\tr_{\rm R}\,L_{\rm R}[\bfe]
= \langle \tr_{\rm R}\,L_{\rm R}[A_0]\rangle$
which underlies most~PNJL/PQM model studies,\footnote{Note that
there are also PNJL/PQM-type model studies which do not use this approximation
but consider an integration over the group SU$(N)$, see e.~g.
Refs.~\cite{Megias:2004hj,Zhang:2010kn}.}
see e.~g. Refs.~\cite{Meisinger:1995ih,Pisarski:2000eq,Mocsy:2003qw,Fukushima:2003fw,%
Ratti:2005jh,Sasaki:2006ww,Schaefer:2007pw,Mizher:2010zb,Skokov:2010wb,Herbst:2010rf,Skokov:2010uh,Nishimura:2009me}.
Although this assumption is convenient and opens up the possibility 
to incorporate lattice results for~$\langle \tr_{\rm R} L_{\rm R}[A_0]\rangle$, 
it may be problematic in quark representations other than the fundamental one
and away from the 
limit of infinitely many colors,~$d({\rm R})\to \infty$, see our discussion below.
For our analytic studies, we shall rather make use of the exact
relations given in Sect.~\ref{sec:GA}. For our numerical evaluation of the quantum 
effective action, we then use the numerical results for~$\langle A_0\rangle$
from a non-perturbative first-principles RG study of the associated order
parameter potential in Polyakov-Landau-DeWitt gauge, see
Refs.~\cite{Braun:2007bx,Braun:2010cy}.
Since it has been found for fundamental matter that PNJL/PQM-type
model studies are sensitive to different parameterizations of the potential for~$\langle \trf L_{\rm F}[A_0]\rangle$, 
see e.~g. Ref.~\cite{Schaefer:2009ui}, it is in fact important to analyze at least some of the
consequences arising from the approximation~$\tr_{\rm R}\,L_{\rm R}[\bfe] = \langle \tr_{\rm R}\,L_{\rm R}[A_0]\rangle$
underlying these model studies.

In general, our ansatz~\eqref{eq:fermionic_action} in the matter sector can be considered as the leading order in a 
systematic derivative expansion. The associated expansion parameter is the anomalous 
dimension~$\eta_{\psi}=-\partial_t \ln Z_{\psi}$ of the quark fields. This ``parameter" is
small as has been found in various previous studies~\cite{Berges:1997eu,Gies:2002hq,Braun:2008pi,Braun:2009si}. 
In fact, it is identical to zero when we consider the four-fermion coupling in the so-called point-like
limit, $\lambda_{\psi}(|p|\ll k)$, see e.~g. Ref.~\cite{Braun:2011pp}. This
is true even if we had allowed for dynamical gauge degrees, provided that one considers
the class of Landau gauges~\cite{Gies:2003dp}, such as Polyakov-Landau-DeWitt gauge. 
As we have discussed above, the latter gauge is implicitly assumed in our work, see Sect.~\ref{sec:GA}.

Apart from an expansion in derivatives, the effective action can be expanded in operators, such as $n$-fermion operators.
Regarding four-fermion operators, we note that our ansatz~\eqref{eq:fermionic_action} for the effective action
is not complete with respect to Fierz transformations even in the limits~$\langle A_0\rangle\to 0$ and~$T\to 0$,
see e.~g. Refs.~\cite{Gies:2003dp,Gies:2005as,Braun:2005uj,Braun:2006jd,Braun:2011pp}. For example,
we have dropped a so-called axial-vector channel interaction which would also contribute to the RG flow
of our coupling~$\bar{\lambda}_{\psi}$ associated with a scalar-pseudoscalar channel. 
From a consideration of a Fierz-complete {basis, however, we only} expect quantitative corrections to our results presented here. 
The main qualitative aspects are expected to persist since
the general structure of the loop integrals remains unchanged~\cite{Braun:2011fw}. For a Fierz-complete study of 
RG flow of four-fermion couplings in QCD, we refer the 
{reader to Refs.~\cite{Gies:2003dp,Gies:2005as,Braun:2005uj,Braun:2006jd,Braun:2009ns,Braun:2011pp}.}
Regarding the role of higher 
fermion operators, e.~g. 8-fermion operators, we note that it can be shown that these operators do not contribute 
to the RG flow of the four-fermion couplings in the point-like limit~\cite{Braun:2011pp}. Beyond the point-like
limit, however, these higher-order operators may very well contribute to the flow of the four-fermion interactions,
see, e.~g., our discussion in Sect.~\ref{sec:PBFP} and
Ref.~\cite{Braun:2011pp}.

In the subsequent section we will show that the purely fermionic formulation of our ansatz~\eqref{eq:fermionic_action}
for the matter sector is convenient for a general discussion of the interplay of the chiral and the
deconfinement phase transition, independent of the fermion representation. In order to compute low-energy
observables, however, a purely fermionic formulation
may not be the first choice since this requires to resolve the momentum-dependence of the fermionic vertices.
In this case, a partially bosonized formulation of our ansatz~\eqref{eq:fermionic_action} might be better suited. Such
a formulation of the effective action can be obtained straightforwardly from a Hubbard-Stratonovich transformation of the underlying
path integral and yields
\be
&&\!\!\!\Gamma_k[\bar{\psi},\psi,\bar{\Phi},\bfe]
= \int d^4 x\,\Big\{ Z_{\psi}\bar{\psi}\left(\I \partial\!\!\!\slash + \bar{g}\gamma_0 \langle A_0\rangle\right)\psi 
\label{Eq:HSTAction} \\
&& \qquad +\,\frac{1}{2}Z_{\Phi}\left(\partial_{\mu}\bar{\Phi}\right)^2
 + {\rm i}\bar{h}\bar{\psi}(\sigma\!+\! {\rm i}\vec{\tau}\cdot\vec{\pi}\gamma_5)\psi
 +\, \frac{1}{2}\bar{m}^2\bar{\Phi}^2 \Big\}\,,\nn
\ee
where the auxiliary bosonic fields~$\bar{\Phi}^{T}=(\sigma,\vec{\pi})$ mediate the interaction
between the fermions. Here, we consider these bosons to be composites of fermions
which do not carry any internal color or flavor charges:  $\sigma\sim (\bar{\psi}\psi)$ and 
$\vec{\pi}\sim (\bar{\psi} \vec{\tau}\gamma_5\psi)$. Chiral symmetry breaking is now signaled
by a non-vanishing expectation value of the $\sigma$ field. Since the mass parameter~$\bar{m}^2$
describes the curvature of the chiral order parameter potential at the origin, the sign of~$\bar{m}^2$
is related to the question whether chiral symmetry is broken in the ground state or not. For~$\bar{m}^2<0$, we
necessarily have $\langle \sigma\rangle \neq 0$. 

In the following we choose the initial conditions for the various couplings in Eq.~\eqref{Eq:HSTAction} such that
\be
\lim_{k\to \Lambda} \bar{m}^2> 0\,,\quad \label{eq:bc01}
\lim_{k\to \Lambda} Z_{\Phi} = 0\,,
\quad
\lim_{k\to \Lambda} Z_{\psi} &=& 1\,.\label{eq:bc2}\nn
\ee
Together with the identity
\be
\bar{\lambda}_{\psi}=\frac{\bar{h}^2}{\bar{m}^2}\,,\label{eq:hstmap}
\ee
the ansatz~\eqref{Eq:HSTAction} can then be mapped onto the 
ansatz~\eqref{eq:fermionic_action} at the initial UV scale $\Lambda$. Thus,
only the ratio of the Yukawa coupling~$\bar{h}$ and the mass parameter~$\bar{m}$ acquires
a physical meaning. In particular, we observe that a large (i.~e. diverging) four-fermion
coupling signals the onset of chiral symmetry breaking, since it can be related to 
a change in the sign of the parameter~$\bar{m}^2$. In fact, the two criteria are equivalent in the
large-$d({\rm R})$ limit due to the absence of fluctuation effects of the Goldstone modes~\cite{Braun:2011pp}.
In any case, we conclude that the fixed-point structure of the coupling~$\bar{\lambda}_{\psi}$ (or, equivalently,
of the couplings~$\bar{m}$ and~$\bar{h}$) is directly linked to the question of chiral symmetry breaking 
in the IR limit. In the following we analyze how this fixed-point structure is related to the order parameter
for center-symmetry breaking, namely~$\tr_{\rm F}L_{\rm F}[\langle
A_0\rangle]$.
This will eventually allow us to 
gain insights into the relation of chiral symmetry breaking and center-symmetry
breaking at finite temperature.
\begin{figure}[t!]
\includegraphics[width=1.0\linewidth]{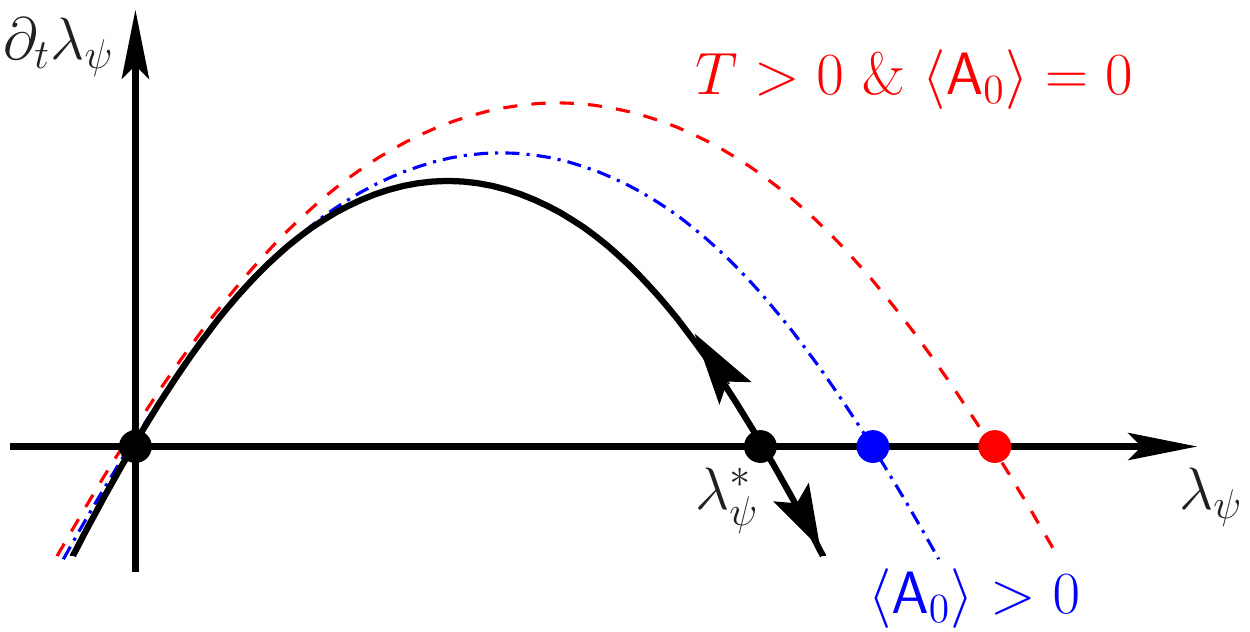}
\caption{
Sketch of the $\beta_{\lambda_{\psi}}$-function of the four-fermion
coupling~$\lambda_{\psi}$ 
for vanishing temperature (black/solid line), finite temperature
and~$\bfe=0$ (red/dashed line), and finite temperature and~$\bfe > 0$
(blue/dashed-dotted line),
see Eq.~\eqref{eq:lpsi_flow}.
The arrows indicate the direction of the RG flow towards the infrared.
The figure has been taken from Ref.~\cite{Braun:2011pp}.}
\label{fig:fp}
\end{figure}
\section{Dynamical Locking Mechanism and the Fermionic Fixed-Point Structure}\label{sec:fermFP}
Let us begin with an analysis of the interplay of center symmetry breaking 
and chiral symmetry breaking using the purely fermionic formulation of the effective action, 
see Eq.~\eqref{eq:fermionic_action}. Our discussion follows closely
the analysis in Ref.~\cite{Braun:2011fw}, where the deformation of the fermionic fixed-point structure due
to the presence of confining dynamics has been analyzed for fundamental quarks in detail.

In our study we consider the value of the background field $\bfe$ as an external input
which is given by the ground state of the corresponding order parameter
potential. As discussed above, the position $\bfe$ of the ground state 
is then directly related to our order parameter 
for center-symmetry breaking, namely ${\rm tr_F} L_{\rm F}[\bfe]$. 
Along the lines of Ref.~\cite{Braun:2011fw}, the RG flow
equation of the four-fermion coupling in the point-like
approximation can be computed for quarks living in any {representation~R}. We find
\be
\beta_{\lambda_{\psi}}\!\equiv\!\partial_t \lambda_{\psi} &=& (2+2\eta_{\psi})\lambda_{\psi} \label{eq:lpsi_flow} \\
&& - \frac{2}{\pi ^2}\Big(2 \!+\! \frac{1}{d({\rm R})}\Big)
\sum_{l=1}^{d({\rm R})} l_{1}^{({\rm F})}(\tau,0,\nu_l^{({\rm R})} |\phi|)\,\lambda_{\psi}^2\,,
\nn
\ee
where the dimensionless renormalized coupling~$\lambda_{\psi}$ is defined as
\be
\lambda_{\psi}=Z_{\psi}^{-2} k^2 \bar{\lambda}_{\psi}\,.
\ee
For example, we have~$d({\rm R=F})=N$ for quarks in the fundamental representation 
and~$d({\rm R=A})=N^2-1$ for quarks in the adjoint representation. For convenience, 
we have introduced the eigenvalues $\nu _l$ of the hermitian matrix given in~Eq.~\eqref{eq:hermmat}:
\be
\nu_l^{({\rm R})} = {\rm spec}\left\{ (T^{a}v^{a})_{ij}\,\; |\; v^2=1 \right\}\,.
\ee
The coupling $\lambda_{\psi}$ depends on the background field $\bfe$ and the dimensionless 
temperature $\tau=T/k$. The threshold function~$l_{1}^{({\rm F})}$ describes 
a regularized one-particle irreducible (1PI)
Feynman diagram with two internal fermion lines. The definition of 
this function can be found in, e.g., Ref.~\cite{Braun:2011fw}.

The RG flow equation~\eqref{eq:lpsi_flow} has two fixed-points:\footnote{At finite temperature, the non-Gau\ss ian
fixed-point is rather a pseudo fixed-point, i.~e. the fixed-point inherits an implicit scale dependence from the dimensionless
temperature~$\tau=T/k$ as well as a dependence on~$\langle A_0\rangle$. Moreover,
the line of pseudo fixed-points~$\lambda_{\psi}^\ast(\tau,\bfe)$ does not
represent a separatrix in the~$(\lambda_{\psi},\tau)$-plane.
However, it represents a strict upper bound for the separatrix in this
plane~\cite{Braun:2011pp}.}
a Gau\ss ian fixed-point ($\lambda_{\psi}\equiv 0$) and a non-trivial
fixed-point~$\lambda_{\psi}^{\ast}(\tau,\bfe)$,
see Fig.~\ref{fig:fp}. The non-Gau\ss ian fixed-point can also be computed
analytically:\footnote{The fixed-point value 
is not a universal quantity as {can be seen} from its
dependence on the regularization scheme. However, 
the statement about the existence of the fixed-point and its qualitative
dependence on the temperature 
and~$\bfe$ is universal.}
\begin{widetext}
\be
\lambda_{\psi}^{\ast}(\tau,\bfe) &=&\left( \frac{1}{ \pi^2}
\left(2\! +\! \frac{1}{d({\rm R})}\right)\sum_{l=1}^{d({\rm R})} l_1^{({\rm F})}(\tau,0,\nu_l^{({\rm R})}|\phi|)  
\right)^{-1}
\nn\\
&=&{\lambda_{\psi}^{\ast}(0,0)} \left(1
+ \frac{1}{d({\rm R})} \sum_{n=1}^{\infty}(-d({\rm R}))^{n}\left[\tr_{\rm R}
(L_{\rm R}[\bfe]^n)
\!+\! \tr_{\rm R} (L_{\rm R}^{\dagger}[\bfe]^n)\right]
\left(
1\!+\!\frac{n}{\tau}
\right){\rm e}^{-\frac{n}{\tau}}
\right)^{-1}\!\!\!,
\label{eq:lfpa0}
\ee
\end{widetext}
where
\be
\lfp\equiv\lambda_{\psi}^{\ast}(0,0) = \frac{6\pi^2}{\left(2d({\rm R})+1\right)}\,.
\ee
Thus, we have~$d({\rm R})\lfp \to {\rm const.}$ for~$d({\rm R})\to \infty$.
Note that we have dropped terms depending on $\eta_{\psi}$ on the right-hand side
of Eq.~\eqref{eq:lfpa0}. As discussed above, this is not an approximation in the point-like limit. 

Before we turn to the case of finite temperature,
we briefly discuss a few basic aspects of the zero-temperature limit. 
In order to solve the RG flow equation~\eqref{eq:lpsi_flow}, we have to choose an initial 
value~$\lambda_{\psi}^{\rm UV}$ at the scale~$k=\Lambda$ for the coupling~$\lambda_{\psi}$.
For~$\luv < \lfp$, we find that the four-fermion coupling approaches the Gau\ss ian fixed-point
in the IR limit, i.~e. the theory becomes non-interacting and chiral symmetry remains intact.
For $\luv > \lfp$, on the other hand, we observe that the four-fermion coupling grows rapidly and
diverges at a finite scale~$\ksb$. This scale signals the onset of chiral symmetry breaking. Below 
this scale, the point-like approximation is no longer justified: the formation of a quark condensate 
and the appearance of Nambu-Goldstone modes requires that we resolve the momentum
dependence of the four-fermion coupling in this regime. For example, this can be done by means of partial bosonization techniques,
see Sect.~\ref{sec:PBFP}. In any case, the chiral symmetry breaking scale~$\ksb$ sets the scale 
for all chiral low-energy observables~$\mathcal O$:
\be
{\mathcal O} \sim \ksb^{d_{\mathcal O}} \sim \left[1 - \left(\frac{\lfp}{\luv}\right)\right]^{\frac{d_{\mathcal O}}{|\Theta|}}\theta (\luv - \lfp)
\label{eq:Ogen}
\,,
\ee
where~$d_{\mathcal O}$ is the canonical mass dimension of the observable~$\mathcal O$ and
the critical exponent~$\Theta$ is defined as
\be
\Theta = - \frac{\partial \beta_{\lambda_{\psi}}}{\partial \lambda_{\psi}}\bigg|_{\lfp}=2\,.
\ee
For details, we refer the reader to Ref.~\cite{Braun:2011pp}. In the following we 
fix~$\luv > \lfp$ at $T=0$. The value of~$\luv$ then determines the symmetry breaking
scale~$\ksb$ and, in turn, the values of the chiral low-energy observables. For our study of 
finite-temperature effects and effects from the confining dynamics parameterized by the
background field~$\bfe$, we leave our choice for~$\luv$ unchanged. This ensures comparability
of our results at zero and finite temperature for a theory defined by a given value of~$\luv$.

At finite temperature and finite~$\bfe$, the fixed-point structure of the theory is deformed compared
to the zero-temperature limit. For illustration purposes, we begin with a 
brief discussion of the case with vanishing gluonic background field. In this case, the pseudo fixed-point
is shifted to larger values at finite temperature, $\lfp (\tau,0) > \lfp$, see Eq.~\eqref{eq:lfpa0}. For a given initial 
value~$\luv > \lfp$, this implies that a critical temperature~$\Tc$ exists above which chiral symmetry
is restored, see Ref.~\cite{Braun:2011pp} for a detailed discussion. Strictly speaking, the
critical temperature~$\Tc$ is defined to be the temperature for which~$1/\lambda_{\psi}\to 0$ 
for~$k\to 0$. From the RG flow equation~\eqref{eq:lpsi_flow}, one then obtains a simple analytic expression
for~$\Tc$:
\be
\Tc = \left(\frac{\Lambda}{\pi}\right)\left[ 1 - \left(\frac{\lfp}{\luv}\right)\right]^{\frac{1}{2}}\theta (\luv - \lfp)\,,
\label{eq:TcAeq0}
\ee
which is accordance with our general statement in Eq.~\eqref{eq:Ogen}.
To derive this expression, we have assumed that $T/\Lambda\ll 1$.

Let us now turn to the case of finite~$\bfe$. For fermions in the fundamental
representation, for example, {we have $\tr _{\rm F}\left( L_{\rm
F}[\bfe]^n\right) \to 0$ in the center symmetric phase for~$n \in {\mathbb N}$ and 
$d({\rm F})=N\to\infty$, see Eq.~\eqref{eq:polnfund}.}
Thus, the temperature-dependent corrections to $\lambda_{\psi}^{\ast}(\tau,\bfe)$ vanish identically 
and we have~$\lambda_{\psi}^{\ast}(\tau,\bfe)\equiv\lambda_{\psi}^{\ast}(0,0)$ for $T\leq \Td$. 
We shall refer to this as a locking mechanism for the chiral phase transition~\cite{Braun:2011fw}. 
Loosely speaking, this means that the chiral phase transition is locked in due to the confining dynamics.
For~$T>\Td$, we have~$\tr_{\rm F}L_{\rm F}[\bfe] >0$ and the fixed-point is again shifted to larger values.
As pointed out in Ref.~\cite{Braun:2011fw}, this implies
that~$\Tc\geq \Td$ in the limit~$\Nc\to\infty$, see 
also Refs.~\cite{Coleman:1980mx,Meisinger:1995ih}. In the case of adjoint
fermions and~$\dr\gg 1$, the temperature-dependent
corrections in Eq.~\eqref{eq:lfpa0} do not vanish identically on all RG scales~$k$ for~$T\leq \Td$
since we now have~$\tr_{\rm A}(L_{\rm A}[\bfe]^n) < 0$ for
these temperatures, see Eq.~\eqref{eq:traconf}. Therefore the (global) sign of
the temperature-dependent corrections changes compared to the case with~$\bfe =
0$. This implies that
the pseudo fixed-point is shifted to smaller values at finite temperature rather than to larger 
values.\footnote{Note that an external magnetic field deforms the fixed-point structure in a similar 
way~\cite{Scherer:2012nn,Fukushima:2012xw}.} For~fixed~$T$ and~$k\to 0$ (i.~e. $\tau\to\infty$), 
{the pseudo fixed-point approaches}
\be
\lambda_{\psi}^{\ast}(\tau\to\infty,\bfe) = \frac{\lfp}{1+\frac{1}{d({\rm A})}}\,.
\ee
For~$d({\rm A})\to \infty$, we have $\lambda_{\psi}^{\ast}(\tau\to\infty,\bfe)\to \lfp$ from below. 
Since $\tr_{\rm A} L_{\rm A}[\bfe]\to 1$ for~$T\gg \Td$, the fixed-point is ``released" and shifted to
larger values. For a given initial value~$\luv>\lfp$, it then follows 
again that~$\Tc \geq \Td$ for~$d({\rm A})\to\infty$.

This analysis can in principle be repeated for any fermion representation,
including fermion representations for which~$\tr_{\rm R} L_{\rm
R}[\bfe]>0$ for~$T<\Td$, such as the ten-dimensional representation, see
Eq.~\eqref{eq:polten}. In the latter case, the temperature-dependent
corrections do not vanish for large values of~$d({\rm R})$. 
However, the finite-temperature shift of the fixed-point is still suppressed by a factor of~$1/d({\rm R})$ compared
to the case with~$\bfe=0$. As a consequence, the chiral phase transition
temperature is increased, but it is not necessarily pushed above the
deconfinement phase transition temperature. Therefore a strict statement about
the relation of the chiral and the deconfinement phase transition 
cannot be made for this class of theories, not even in the large-$\dr$ limit.

From Eq.~\eqref{eq:lpsi_flow}, we can derive an implicit equation for the chiral phase transition temperature~$\Tc$ 
provided that we neglect a possible RG scale {dependence of~$\tr_{\rm R}L_{\rm R}[\bfe]$}:
\be
\Tc^2 = \frac{1}{{\mathcal P}_{\rm R}(\Tc)}\left(\frac{\Lambda}{\pi}\right)^2 \left[ 1\! -\! \left(\frac{\lfp}{\luv}\right)\right] \theta (\luv\! -\! \lfp),
\label{eq:TcFourFPol}
\ee
where
{
\be
{\mathcal P}_{\rm R}(T)&=& -\frac{6}{d({\rm R})\pi^2}\sum_{n=1}^{\infty}\frac{1}{n^2} (-d({\rm R}))^n
\big[ 
\tr_{\rm R} (L_{\rm R}[\bfe]^n) \nn\\
&& \qquad\qquad\qquad\qquad\quad + \tr_{\rm R} (L_{\rm
R}^{\dagger}[\bfe]^n)\big]\,.
\label{eq:TcFourFPolPDEF}
\ee}
In order to derive this equation, we have again assumed that~$T/\Lambda\ll 1$.
For~$\bfe \to 0$, we have~{${\mathcal P}_{\rm R}(T)=1$}, as 
expected from Eq.~\eqref{eq:TcAeq0}.
Since~$\tr_{\rm R}L_{\rm R}[\bfe]$ depends on~$\Td$, Eq.~\eqref{eq:TcFourFPol} relates the 
chiral phase transition temperature~$\Tc$ to the deconfinement phase transition temperature~$\Td$.

For fermions in the fundamental representation and~$N\to\infty$,
we have~${\mathcal P}_{\rm F}(T)=0$ for~$T\leq \Td$ and~${\mathcal P}_{\rm F}(T)>0$ for $T>\Td$. 
Thus, no finite solution of Eq.~\eqref{eq:TcFourFPol} exists for~$\Tc <\Td$.
In accordance with our fixed-point analysis, we conclude
that~$\Tc \geq \Td$. 

For fermions in the adjoint representation and~$d({\rm A})\gg 1$, we 
have~${\mathcal P}_{\rm A}(T) \approx -1/d({\rm A})$ for $T\leq \Td$. As in the case of fundamental matter fields,
this implies again that~$\Tc \geq \Td$. For fermion representations with~$0<{\mathcal P}_{\rm R}(T)<1$
for~$T<\Td$, we observe that the chiral phase transition temperature is still shifted to larger
{values compared to the case with~$\bfe =0$.}
However, a definite statement about the temperature-order of the two phase transitions 
cannot be made in this case.
\begin{figure*}[t]
\includegraphics[width=0.47\linewidth]{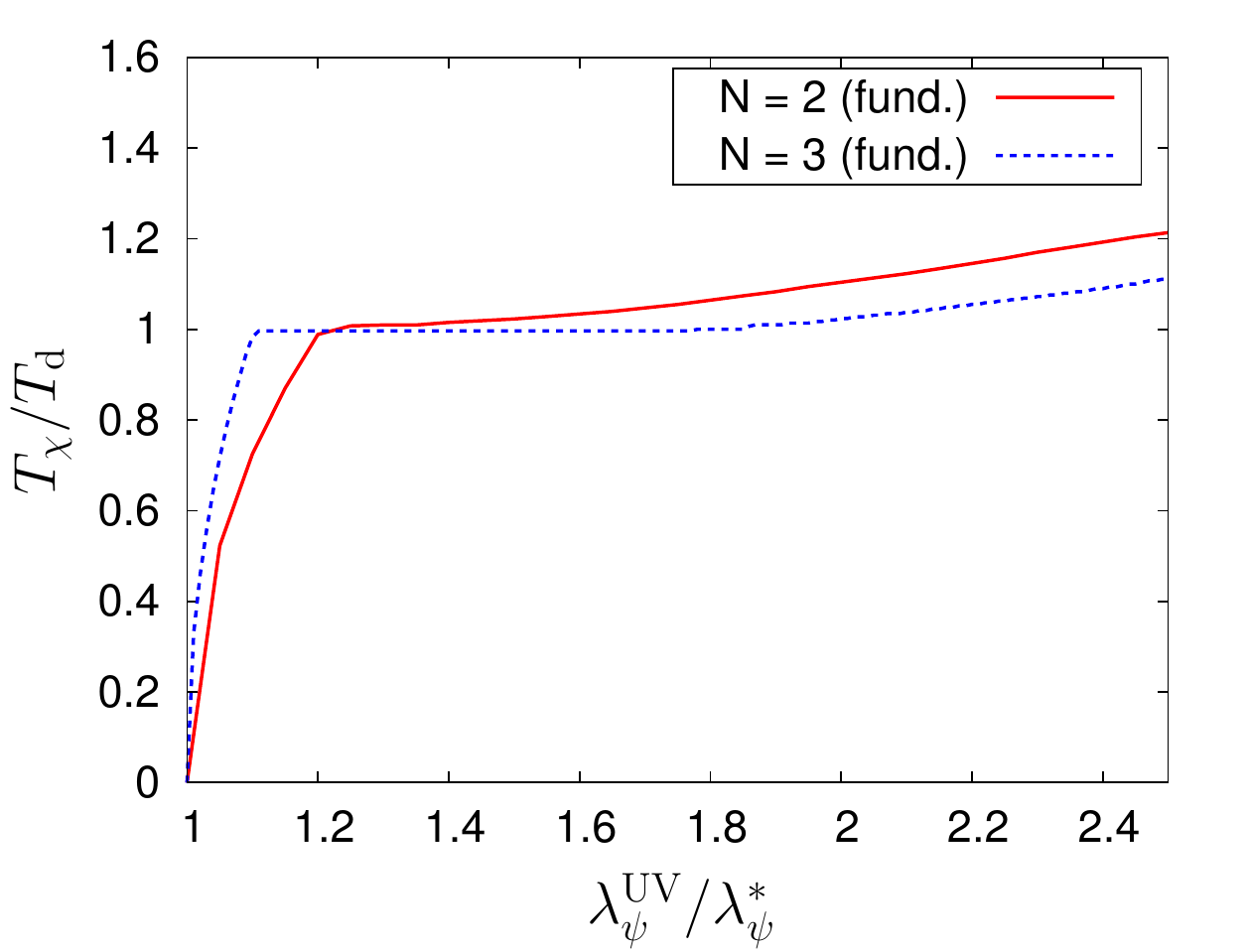}
  \hspace*{0.8cm}
\includegraphics[width=0.47\linewidth]{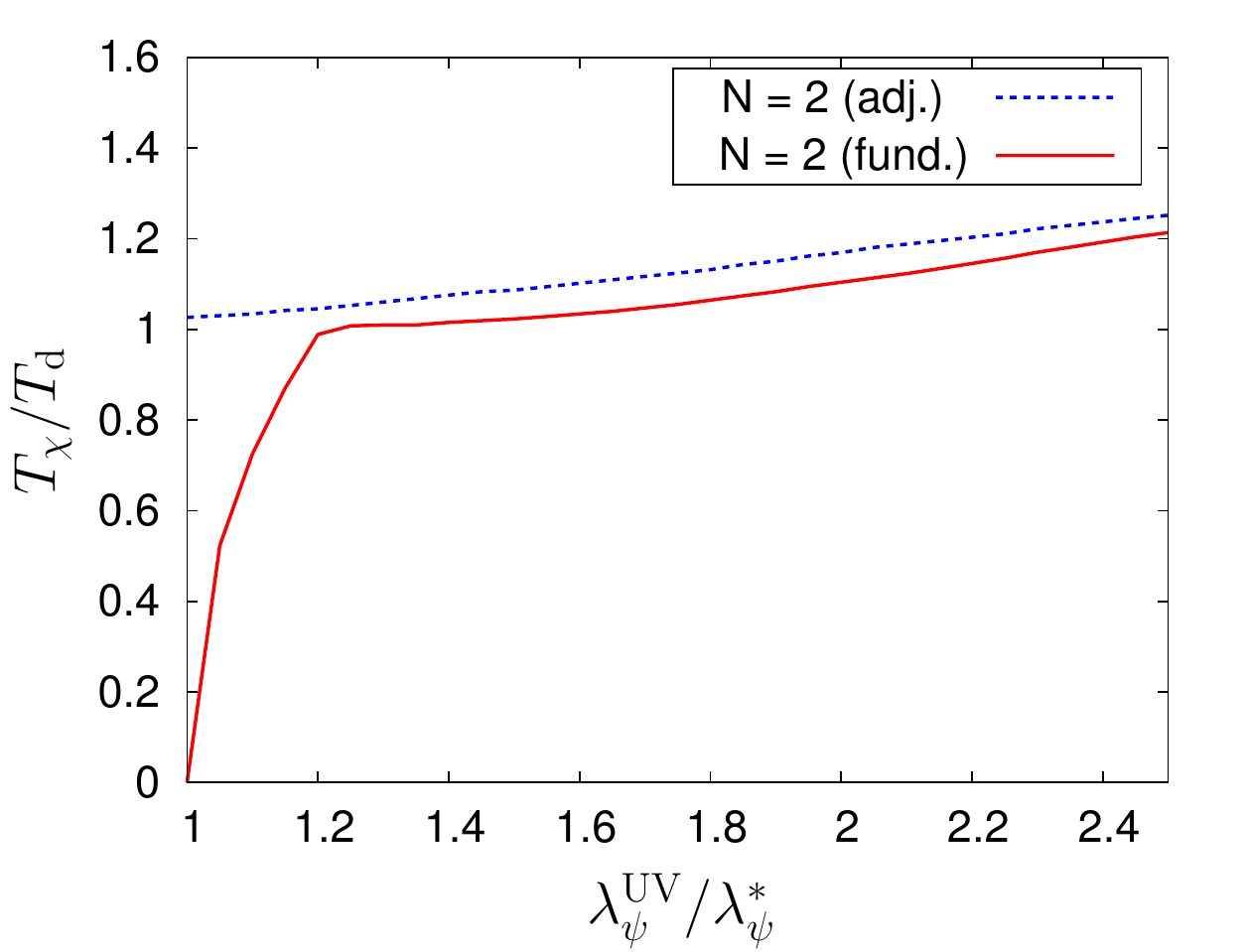} 
\caption{Left panel: Phase diagram  in the plane spanned by the
temperature and the rescaled coupling $\luv/\lfp$ for~$\Nf=2$ massless quark flavors 
in the fundamental representation and~$N=2$ colors (red/solid line) as well as for $N=3$
colors (blue/dashed line), see also Ref.~\cite{Braun:2011fw}.
Note that there is no splitting of the phase boundary (i.~e. $\Tc\simeq \Td$) 
for small $\luv$ in the large-$N$ limit, see Eq.~\eqref{eq:TcFourFPol} and discussion thereof.
Right panel: $\Tc/\Td$ as a function of~$\luv/\lfp$ for~$N_{\rm f}=2$
massless quarks in the fundamental representation ($N=2$) (red/solid line) as well as for quarks in the
adjoint representation (blue/dashed line).
}
\label{fig:TlambdaF}
\end{figure*}

Let us now turn to the case of finite values of~$d({\rm R})$. For fermions in the fundamental representation, we then
find~${\mathcal P}_{\rm F}(T)=1/N^2 >0$ for~$T\leq\Td$. For~$T>\Td$, ${\mathcal P}_{\rm F}(T)$ increases monotonically
from~${\mathcal P}_{\rm F}(\Td)=1/N^2$ to ${\mathcal P}_{\rm F}(T\to\infty)\to 1$.
In fact, right above the deconfinement phase transition, the quantity ${\mathcal P}_{\rm F}(T)$ increases rapidly
since~$\tr_{\rm F}L_{\rm F}[\bfe]$ increases rapidly.
Since~${\mathcal P}_{\rm F}(T)$ is finite for all temperatures, we find a regime
where~$\Tc < \Td$ for~$\luv/\lfp\gtrsim 1$. 
In the left panel of Fig.~\ref{fig:TlambdaF}, we show our numerical results for~$\Tc/\Td$ as a function 
of~$\luv/\lfp$. For larger values of~$\luv/\lfp$, a window in parameter space
opens up in which
the chiral and the deconfinement phase transition (almost) coincide. In the limit~$N\to\infty$, this locking
window extends down to~$\luv/\lfp=1$, as illustrated by a comparison of our results for~$N=2$ and~$N=3$
in Fig.~\ref{fig:TlambdaF}. Note that the locking window for~$\luv/\lfp$ can be related to a locking window for 
low-energy observables, such as the pion decay constant. We shall come back to this in Sect.~\ref{sec:PBFP}.

In case of adjoint fermions and finite~$d({\rm A})$, we have ${\mathcal P}_{\rm A}(T) \leq 0$ for $T\leq \Td$. For example,
we have~${\mathcal P}_{\rm A}(T)=-1$ for $N=2$. For~$T\gtrsim\Td$, ${\mathcal P}_{\rm A}(T)$ increases rapidly
and changes its sign. For~$T\gg \Td$, it then
approaches~${\mathcal P}_{\rm A}(T)=1$. Since we have~${\mathcal P}_{\rm A}(T)
\leq 0$ even for finite~$N$, we find that the chiral phase transition
temperature is larger than the deconfinement phase transition temperature, 
independent of our choice for~$\luv/\lfp > 1$ and~$N\geq 2$, 
see right panel of Fig.~\ref{fig:TlambdaF} for our numerical results for~$N=2$.
Note that this observation is compatible
with lattice results of SU($2$) gauge theory with two adjoint quarks, 
see Refs.~\cite{Karsch:1998qj,Engels:2005te,Bilgici:2009jy}. In our analysis, 
it can be traced back to the deformation of the fermionic
fixed-point structure in the presence of gauge dynamics.

To obtain the numerical results in Fig.~\ref{fig:TlambdaF}, we have employed data for~$\bfe(T)$ as obtained from
an RG study of the associated order parameter potential for~SU($2$) and~SU($3$) Yang-Mills 
theory~\cite{Braun:2007bx,Braun:2010cy}. However, we did not take into account the back-reaction of
the matter fields on the order parameter potential associated with~$\bfe$. In the case of fundamental matter, 
we expect that this back-reaction will shrink the size of the locking window since it further increases the 
quantity~${\mathcal P}_{\rm F}(T)$ at low temperatures. For adjoint quarks, the back-reaction will also 
increase~${\mathcal P}_{\rm A}(T)$. Nevertheless, it may remain negative over a wide range of temperatures.
Therefore we may still have~$\Tc > \Td$ for all values of~$\luv/\lfp > 1$, at least for~$N=2$.

Let us add a word of caution on the treatment of the quantity~$\tr_{\rm R}L_{\rm R}[\bfe]$
in standard PNJL/PQM model approaches. In these studies,
one relies on the assumption that~$\tr_{\rm R}L_{\rm R}[\bfe]=\langle \tr_{\rm R}L_{\rm R}[A_0]\rangle$. 
For~$\langle \tr_{\rm R}L_{\rm R}[A_0]\rangle$,
one then uses lattice data as input. Whereas such an approach would lead to similar 
conclusions for fundamental quarks ($\langle \tr_{\rm F}L_{\rm F}[A_0]\rangle\geq 0$ and $\tr_{\rm F}L_{\rm F}[\bfe]\geq 0$), 
the situation is different for adjoint quarks. In the latter case, we have $\langle \tr_{\rm A}L_{\rm A}[A_0]\rangle>0$ 
but $\tr_{\rm A}L_{\rm A}[\bfe]$ can assume both positive and negative values as discussed above.

Before we now enter the discussion of the RG flows of the partially bosonized formulation of the matter sector, we would like to 
comment on the number of parameters in our study. Up to this point, 
our discussion suggests that our study only relies
on a single parameter in the matter sector apart from the UV cutoff~$\Lambda$, 
namely on the initial value $\luv$. Strictly speaking, however, the non-trivial fixed-point of the four-fermion
interaction is an artifact of our point-like approximation. With the aid of the partially bosonized formulation, we will resolve part of
the momentum dependence of the four-fermion interaction. We will then find that the matter sector depends on three parameters: 
the Yukawa coupling~$\bar{h}$, 
the bosonic mass parameter~$\bar{m}$ and the UV cutoff~$\Lambda$,
see Eq.~\eqref{Eq:HSTAction}. This is a substantial difference to, e.~g., fermion models in $d<4$ space-time dimensions,
where we only have a single parameter in both formulations, see e.~g. Ref.~\cite{Braun:2010tt}.
There, the non-trivial fixed-point 
of the four-fermion coupling can be mapped onto a corresponding non-trivial fixed-point
in the plane spanned by the renormalized Yukawa coupling~$h$
and the dimensionless renormalized bosonic mass parameter~$m$.
In our case, the role of the 
non-trivial fixed-point in the purely fermionic formulation is taken over by a separatrix in the $(h^2,m^2)$-plane
in the partially bosonized formulation. The shift of the non-trivial fixed-point of the four-fermion coupling due 
to the gauge dynamics then turns into a corresponding shift of this separatrix. 
The mapping between the two formulations is discussed in detail in the subsequent section. 
Being aware of this subtlety, the discussion of the fermionic fixed-point structure is still useful and
nicely illustrates the mechanism underlying the interplay of the chiral and the deconfinement phase transition.

\section{Partial Bosonization and the large-$\dr$ Expansion}\label{sec:PBFP}
\subsection{Gap Equation}\label{sec:LNPBFPgap}
In this subsection, we briefly discuss how our study of fermionic RG flows is related to the 
gap equation for the fermion mass in the large-$\dr$
limit. For related QCD reviews on Dyson-Schwinger equations, we refer the reader
to Refs.~\cite{Alkofer:2000wg,Fischer:2006ub,Maas:2011se,Roberts:2012sv}.

Starting from the partially bosonized action given in Eq.~\eqref{Eq:HSTAction}, we can derive the gap
equation for the vacuum expectation value~$\bar{\Phi}_0=(\langle \sigma\rangle,\vec{0})$ 
of the scalar fields and the fermion mass, respectively.
To this end, we first consider the so-called classical action~$S\simeq \Gamma_{k\to\Lambda}$ which appears 
in the functional integral. Since the fermions appear only as bilinears in the action~$S$, these fields
can be integrated out straightforwardly. From the resulting expression, we obtain the (fully) bosonized effective 
action~$\Gamma_{\rm B}[\sigma,\vec{\pi}]$, which is a highly non-local object. From the stationary condition,
\be
\frac{\delta \Gamma_{\rm B}[\sigma,\vec{\pi}]}{\delta \sigma}\bigg|_{\bar{\Phi}_0}=0\,,
\ee
we then find the gap equation for~$\langle \sigma\rangle$:\footnote{Here and in the following we assume 
that the ground state is homogeneous.}
\be
1 &=& 8 \left(\frac{\bar{h}^2_{\Lambda}}{\bar{m}^2_{\Lambda}}\right) 
{\rm Tr}
\int \!\frac{d^3 p}{(2\pi)^3}
\left[  
G_{\psi}^{(n)}(\vec{p}^{\,2},{\bar{g}\langle A_0\rangle},\langle\sigma\rangle)
\right. \nn\\
&& \qquad\; \left. 
- G_{\psi}^{(n)}(\Lambda^2,{\bar{g}\langle A_0\rangle},\langle\sigma\rangle)
\right]\theta(\Lambda^2 -\vec{p}^{\,2})\,, 
\label{eq:gapsigma0}
\ee
%
where
\be
\!\!\! G_{\psi}^{(n)}(\vec{p}^{\,2},{\bar{g}\langle A_0\rangle},\langle\sigma\rangle)=
\frac{1}{ (\nu _n \!+\! {\bar{g}\langle A_0\rangle}) ^2\! 
+\! \vec{p}^{\,2}\! +\! {\bar h^2}\langle \sigma \rangle^2}\,
\ee
and~$\nu_n=(2n+1)\pi T$. The trace~${\rm Tr}$  is {defined as follows:
\be
{\rm Tr} \cdots =\tr _{\rm R}\,T\!\sum_{n=-\infty}^{\infty} \cdots\,.
\ee
In Eq.~\eqref{eq:gapsigma0} we} have dropped the trivial solution~$\langle \sigma\rangle =0$.
The integral on the right-hand side of the gap equation 
represents a Feynman integral with two internal fermion lines
and two (Yukawa) vertices. In order to regularize this integral, we have employed
a regularization scheme\footnote{Often, a sharp cutoff is
used to regularize the gap equation. In this case, the $\Lambda$-dependent term
in the square brackets in Eq.~\eqref{eq:gapsigma0} is absent, see Ref.~\cite{Braun:2011pp}.} 
which corresponds to the one used to derive the RG flow equation for the four-fermion coupling~$\bar{\lambda}_{\psi}$ in the previous section,
see Eq.~\eqref{eq:lpsi_flow}.
The structure of the loop integral in the gap equation and
on the right-hand side of the flow equation~\eqref{eq:lpsi_flow} is indeed 
identical.\footnote{Recall also that~$\bar{h}^2_{\Lambda}/\bar{m}^2_{\Lambda}$ can 
be identified with $\bar{\lambda}_{\psi}^{\rm UV}\equiv\bar{\lambda}_{\psi,\Lambda}$.} 
However, the prefactor on the right-hand side of the gap equation is only correct in leading order in
the large-$d({\rm R})$ expansion,
in contrast to the associated prefactor in the flow equation~\eqref{eq:lpsi_flow} of the four-fermion coupling.\footnote{Loosely speaking, the 
trace~$\tr_{\rm R}$ yields a factor of~$d({\rm R})$.} 
Our general arguments concerning the relation of the chiral and the deconfinement phase transition
in the previous section are not affected by this prefactor. {The latter plays only a
qualitative, but no quantitative role. 
Therefore our findings concerning} the interplay
of the chiral and the deconfinement phase transition can be obtained from the gap equation~\eqref{eq:gapsigma0} 
as well, as it should be. In fact, the fixed-points of the (dimensionless) four-fermion coupling can be viewed as critical 
values for the dimensionless quantity~$\Lambda^2\bar{h}^2_{\Lambda}/\bar{m}^2_{\Lambda}$. This follows also from
our discussion below Eq.~\eqref{Eq:HSTAction}. We refrain here from discussing
this further. For a detailed discussion of fermionic
RG flows and their relation to the gap equation, we refer the reader to, e.~g., Ref.~\cite{Braun:2011pp}.

\subsection{RG Flow at Large $\dr$}\label{sec:LNPBFP}

Let us now discuss the fixed-point structure and the locking mechanism in the 
partially bosonized formulation of the matter sector, see Eq.~\eqref{Eq:HSTAction}.
This formulation has the advantage that it allows us to systematically 
resolve the momentum dependence of the four-fermion interaction by
means of a derivative expansion, see e.~g. Ref.~\cite{Braun:2011pp}.
Eventually, this allows us to relate the initial value~$\luv$ of the
four-fermion coupling to physical low-energy observables, such as meson masses and the
pion decay constant~$f_{\pi}$. The phase diagrams in the~$(T,\luv)$ plane can then be
translated into phase diagrams in, e.~g., the $(T,f_{\pi})$ plane. In other words, the partially bosonized
formulation gives us access to the phase with broken chiral symmetry in the ground state.

From the effective action~\eqref{Eq:HSTAction}, we obtain the RG flow equations for the
partially bosonized formulation. In leading order of an expansion in powers of~$d({\rm R})$, 
we find the following equations for the chirally symmetric regime:
\be
\eta_{\Phi}&=& \frac{2}{3\pi^2}   \sum_{l=1}^{d({\rm R})}{\mathcal M}_{4,\perp}^{({\rm F})}(\tau,0,\nu_l |\phi|) h^2 \,,
\label{eq:etaphiLN}
\\
\eta_{\psi}&=& 0\,,
\\
\partial_t h^2 &=& (2\eta_{\psi} + \eta_{\Phi})h^2\,,\label{eq:h2LN}
\ee
\be
\partial_t m^2 &=& (\eta_{\Phi}\!-\! 2)m^2 +\frac{4}{\pi ^2}
\sum_{l=1}^{d({\rm R})} l_{1}^{\rm (F)}(\tau,0,\nu_l |\phi|) h^2,\label{eq:m2flowLN}
\ee
\be
\partial _t \lambda_{\Phi}&=& 2 \eta_{\Phi}\lambda_{\Phi}  -\,\frac{8}{\pi ^2}
\sum_{l=1}^{d({\rm R})} l_{2}^{\rm (F)}(\tau,0,\nu_l |\phi|) h^4\,,\label{eq:lflow}
\ee
where~$\eta_{\Phi}=-\partial_t \ln Z_{\Phi}$, $\eta_{\psi}=-\partial_t \ln Z_{\psi}$, 
$h^2=Z_{\Phi}^{-1}Z_{\psi}^{-2}\bar{h}^2$,~$m^2=k^{-2}Z_{\Phi}^{-1}\bar{m}^2$,
and~$\lambda_{\Phi}=Z_{\Phi}^{-2}\bar{\lambda}_{\Phi}$.
The threshold functions can be found in App.~\ref{app:thresfcts} and~Ref.~\cite{Braun:2008pi}.
We add that we do not distinguish between wave-function renormalizations longitudinal ($Z_{\psi}^{\|}$, $Z_{\Phi}^{\|}$)
and transversal  ($Z_{\psi}^{\perp}$, $Z_{\Phi}^{\perp}$) to the heat-bath. In the following we identify the 
corresponding wave-function renormalizations, $Z_{\psi}^{\|}=Z_{\psi}^{\perp}\equiv Z_{\psi}$ 
and~$Z_{\Phi}^{\|}=Z_{\Phi}^{\perp}\equiv Z_{\Phi}$. In Ref.~\cite{Braun:2009si}, it has indeed been found for the case~$\bfe=0$
that the difference is small at low temperatures and only yields mild corrections to, e.~g., the thermal masses 
close to and above the chiral phase transition. In our study of the partially bosonized formulation, we also include the
RG flow of the four-boson coupling~$\bar{\lambda}_{\Phi}$ associated with an additional term~$\sim (\bar{\lambda}_{\Phi}/8)\bar{\Phi}^4$
in our ansatz~\eqref{Eq:HSTAction}. Since we have~$\bar{\Phi}\sim (\bar{\psi}{\mathcal O}\psi)$, 
this type of interaction parametrizes higher-order fermionic self-interaction terms. These interactions 
are generated dynamically in the RG flow due to Yukawa-type quark-meson interactions.
The initial value of the associated coupling is set to zero in our studies, i.~e.~$\bar{\lambda}_{\Phi}=0$ at~$k=\Lambda$. This allows
us to map the partially bosonized theory onto our purely fermionic ansatz for the matter sector at the initial RG scale~$\Lambda$.

{In the large-$d({\rm R})$ limit, }
the flow of the four-boson coupling (and also of higher bosonic self-interactions~$\sim \bar{\Phi}^{2n}$) does 
not contribute to the RG flows of~$Z_{\Phi}$, $Z_{\psi}$, $h$ and~$m$, at least in the chirally symmetric regime. 
This corresponds to the fact that the RG flow of the four-fermion coupling is decoupled
from the RG flow of fermionic $n$-point functions of higher order, such as $8$-fermion interactions. 

Since we consider the large-$d({\rm R})$ limit in this subsection, we only have purely fermionic
loops appearing on the right-hand side of the flow equations. 
We would like to stress that the large-$d({\rm R})$ expansion should not
be confused with the widely used local potential approximation (LPA) where 
the running of the wave-function renormalizations is not taken into account.

Let us now relate the RG flow of the partially bosonized formulation to the RG flow of the purely fermionic 
formulation. Using Eq.~\eqref{eq:hstmap} together with the flow equations~\eqref{eq:h2LN} and~\eqref{eq:m2flowLN},
we recover the flow equation~\eqref{eq:lpsi_flow} of the four-fermion coupling~$\lambda_{\psi}$ in the large-$d({\rm R})$ limit, i.~e.
\be
\partial _t \left(\frac{h^2}{m^2}\right)\Bigg|_{\dr\to\infty} \equiv \partial _t \lambda_{\psi}\Bigg|_{\dr\to\infty}\,,
\label{eq:h2m2LN}
\ee
see also Eq.~\eqref{eq:lambdapsiBLN}. 
Thus, only our choice for the ratio~$h^2/m^2$ at the initial RG scale determines
whether chiral symmetry is broken in the IR limit~($k\to 0$). Since the flow equation for the ratio~$h^2/m^2$ is identical
to the one for~$\lambda_{\psi}$, {we already anticipate that our} statements concerning the
temperature-order of the chiral and
the deconfinement phase transition still hold in the partially bosonized 
formulation, see also Ref.~\cite{Braun:2011fw}.
\begin{figure}[t]
\includegraphics[width=1.0\linewidth]{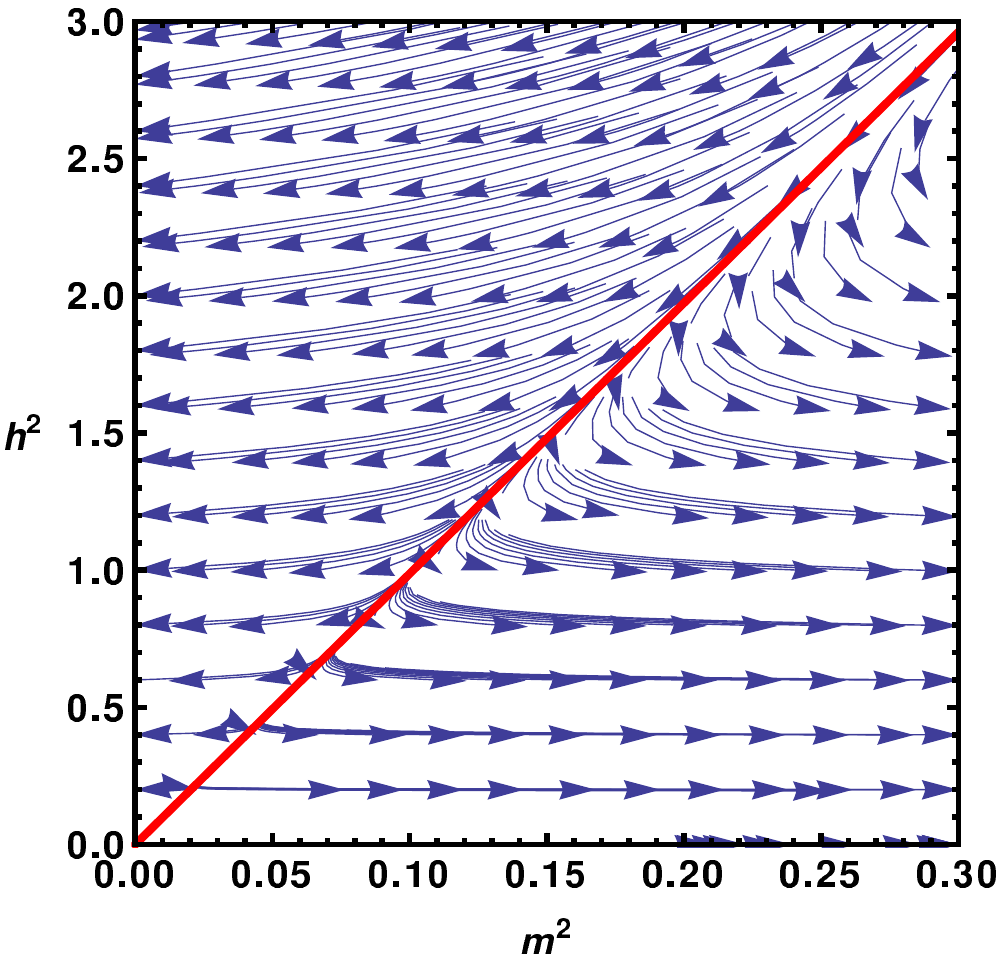}
\caption{
RG flow for fundamental fermions
and~$N=3$ in leading order in the $1/d(R)$-expansion in the $(h^2,m^2)$-plane (at zero temperature).
The red (solid) line represents the separatrix (critical manifold). 
The arrows indicate the direction of the RG flow towards the infrared, see text for an interpretation.
}
\label{fig:sepLN}
\end{figure}

After having shown the equivalence of the RG flow of~$h^2/m^2$ and~$\lambda_{\psi}$ in the chirally
symmetric regime, we now discuss the number of parameters in the matter sector of our model.
Relation~\eqref{eq:h2m2LN} seems to suggest that we only have one parameter,
namely the ratio~$h^2/m^2$ at the initial RG scale. Indeed, the value of the symmetry breaking scale~$\ksb$ depends
only on our choice for~$h^2/m^2$ at the initial RG scale. This suggests that a non-trivial IR repulsive fixed-point
also exists in the plane spanned by the couplings~$h^2$ and~$m^2$. From the above set of flow equations, however, 
we read off that the system has only a Gau\ss ian (non-interacting) 
fixed-point $(h^2_{\ast},m^2_{\ast})_{{\rm Gau\ss}}=(0,0)$, but no non-Gau\ss ian
fixed-point. This seems to be in contradiction to Eq.~\eqref{eq:h2m2LN} and to our results 
{from the purely fermionic formulation.} 
Apart from the Gau\ss ian fixed-point, we also observe that a separatrix exists
in the plane spanned by~$h^2$ and~$m^2$. The latter
separates the $(h^2,m^2)$-plane into two disjunct regimes, see Fig.~\ref{fig:sepLN}. 
In the large-$d({\rm R})$ limit, 
the functional form of the separatrix can be computed analytically. At~$T=0$, {we find
\be
 h^2_{\rm sep.}(m^2)  = \frac{3\pi^2 m^2}{d({\rm R})} \equiv  \lambda_{\psi,\infty}^{\ast} m^2 \,,
\ee
where~$\lambda_{\psi,\infty}^{\ast}$ is the value of the fixed point~$\lambda_{\psi}^{\ast}$ in the 
large-$\dr$ limit.
Choosing} initial conditions~$(h^2_{\Lambda},m^2_{\Lambda})$ in the domain to the left of the separatrix, we find that 
the system flows into the regime with~$m^2<0$, in which chiral symmetry is broken in the ground state.
On the other hand, the system remains in the chirally symmetric regime, if we initialize the flow
in the domain to the right of the separatrix, see  Fig.~\ref{fig:sepLN}. Loosely speaking, we have found that 
the separatrix takes over the role of the non-Gau\ss ian 
fixed-point~$\lfp$ which is present in the point-like approximation of the purely fermionic formulation. 

To further clarify the fate of the seemingly missing non-trivial fixed-point in the $(h^2,m^2)$-plane, we briefly
consider the case~$2<d<4$. In this case, a non-trivial
fixed-point indeed exists in NJL-type and Gross-Neveu-type models, see e.~g. Ref.~\cite{Braun:2010tt}.
This follows immediately from a consideration of the RG flow of the dimensionless renormalized Yukawa 
coupling~$h^2=k^{d-4}Z_{\Phi}^{-1}Z_{\psi}^{-2}\bar{h}^2$:
\be
\partial _t h^2 = (d-4 + \eta_{\Phi} +2\eta_{\psi})h^2\,.
\ee
This differential equation has a Gau\ss ian fixed-point and a non-Gau\ss ian fixed-point~$h^2_{\ast}$ for~$2<d<4$
since~$\eta_{\Phi} \sim h^2$ and~$\eta_{\Phi} >0$, see Refs.~\cite{Braun:2010tt}. In the
$(h^2,m^2)$-plane, we therefore have a non-trivial fixed-point with an IR attractive and IR repulsive direction
for~$2<d<4$. This non-trivial fixed-point represents the intersection point of two separatrices in the $(h^2,m^2)$-plane
and corresponds to the non-trivial fixed-point of the associated four-fermion coupling. For~$d\to 4$, this fixed-point
then merges with the Gau\ss ian fixed-point. 

The non-existence of the non-Gau\ss ian fixed-point in~\mbox{$d=4$} implies 
that the Yukawa coupling $h$ and the UV cutoff~$\Lambda$ should be considered as
parameters of the theory, in addition to the
ratio~$h^2_{\Lambda}/m^2_{\Lambda}$. In fact, for any finite UV
cutoff, we can still define a critical value for the ratio
$h^2/m^2\sim\lambda_{\psi}$ above which
spontaneous chiral symmetry breaking occurs in the long-range limit~$(k\to 0)$. 
However, since no non-trivial fixed-point with an IR attractive direction exists in the $(h^2,m^2)$-plane, the value of the Yukawa coupling at the
symmetry breaking scale~$\ksb$ depends (strongly) on the initial conditions for the bosonic mass parameter and the 
Yukawa coupling itself. 
Once the system enters the regime with~$m^2<0$, the RG flow of the Yukawa coupling
effectively ``freezes". In this low-energy
regime, the fermions acquire a mass and fermionic loops are therefore generically suppressed, see also discussion below. 
Hence we have three parameters in our simplified model ansatz for the matter sector, 
namely~$h^2_{\Lambda}/m^2_{\Lambda}$, $h^2_{\Lambda}$ and~$\Lambda$.
We would like to emphasize that there is indeed only a single parameter in~$2<d<4$, as discussed in detail for the Gross-Neveu model
in Ref.~\cite{Braun:2010tt}.
\begin{figure}[t]
\includegraphics[width=0.75\linewidth]{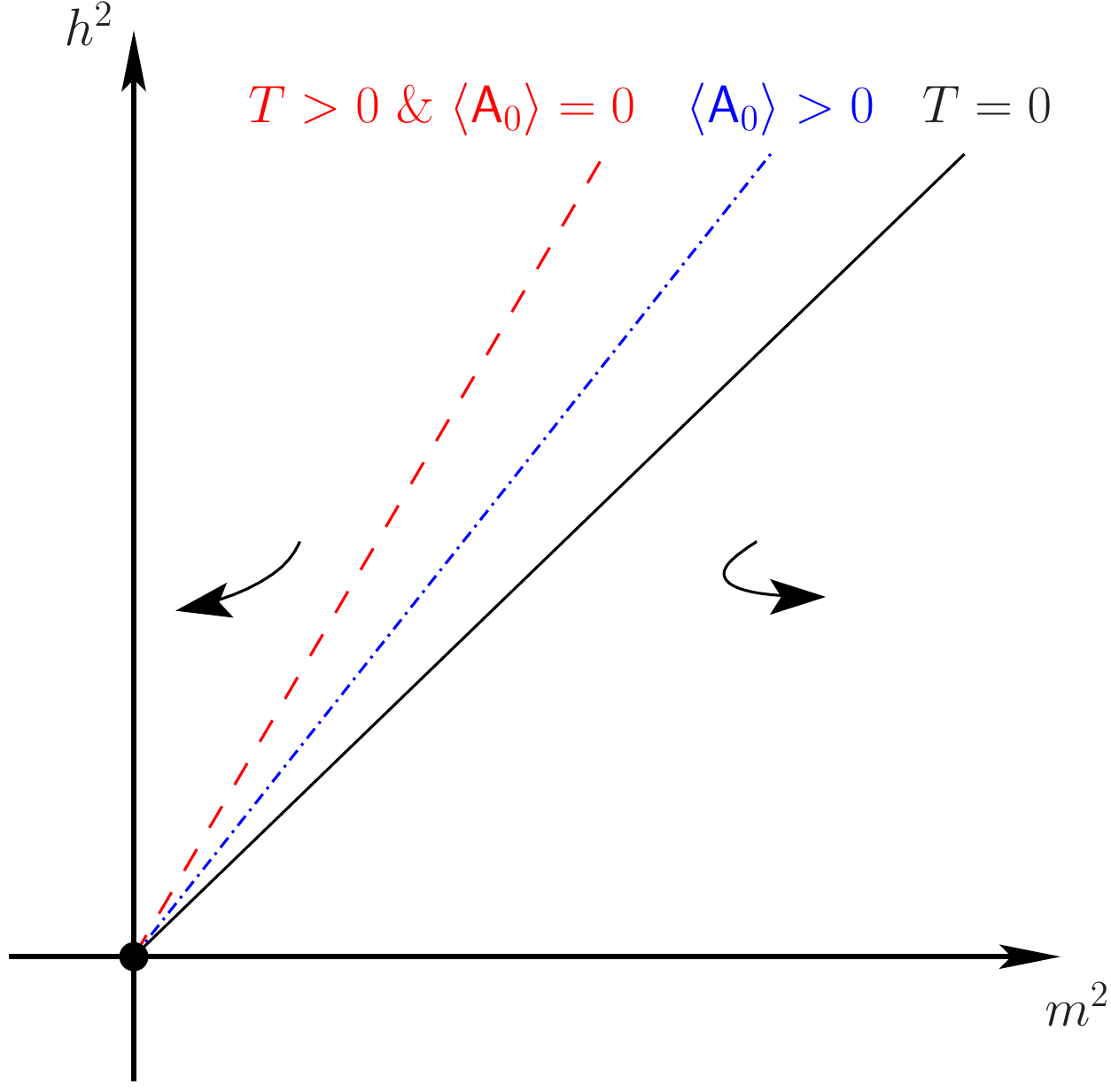}
\caption{
Sketch of the RG flow in leading order in the $1/\dr$-expansion in the $(h^2,m^2)$-plane.
The black dot denotes the Gau\ss ian fixed-point. The separatrices are sketched for three different
cases: vanishing temperature (black line), finite temperature
and~$\bfe=0$ (red/dashed line), and finite temperature and~$\bfe > 0$ (blue/dashed-dotted line).
The dependence of the separatrices on the temperature and~$\bfe$ reflects the behavior of the
non-Gau\ss ian fixed-point of the four-fermion coupling, see Fig.~\ref{fig:fp}.
The arrows to the left and to the right of the 
separatrices indicate the direction of the RG flow towards the infrared, respectively.
}
\label{fig:h2m2flow}
\end{figure}

Let us now analyze the dynamics at finite~$T$ and~$\bfe$. Our discussion of Eq.~\eqref{eq:h2m2LN} and of the
RG flow in the $(h^2,m^2)$-plane at zero temperature already suggests that our general arguments 
concerning the relation
of the deconfinement and the chiral phase transition in Sect.~\ref{sec:fermFP} are still valid. 
This is not too surprising: the point-like approximation in the purely fermionic formulation 
is a reasonable approximation in the chirally symmetric regime where the bosonic mass parameter~$m^2$ is large
over a wide range of scales and therefore suppresses the non-trivial momentum-dependence of the vertices.\footnote{Recall
that the bosons mediate the interaction between the fermions in the partially bosonized formulation. In this spirit,
the boson propagators parametrize the momentum dependence of the four-fermion coupling, see e.~g. 
Ref.~\cite{Braun:2011pp} for a detailed discussion.}
In any case,
we now have to study the behavior of the separatrix in the $(h^2,m^2)$-plane for finite temperature~$T$ and
finite~$\bfe$. To this end, we may even
consider the dimensionless temperature~$\tau=T/k$ as an additional coupling of the theory. Thus, the separatrix is no longer
a one-dimensional manifold as it is the case at zero temperature. It rather represents a two-dimensional manifold.
Again, the functional form of this critical manifold can be computed analytically. For~$\tau = T/\Lambda \ll 1$, {we find
\be
h^2_{\rm sep.}(m^2,\tau)=\frac{\lambda_{\psi,\infty}^{\ast} m^2}{1-\pi^2 {\mathcal P}_{\rm R}(T) \tau^2 }\,,\label{eq:critmanifold}
\ee
where~${\mathcal P}_{\rm R}(T)$ is} defined in Eq.~\eqref{eq:TcFourFPolPDEF}.
We observe that the shape of the critical manifold depends on the temperature and the 
order parameter for center symmetry breaking, see Fig.~\ref{fig:h2m2flow}.

The critical manifold allows us to define a necessary condition for chiral symmetry breaking at finite temperature.
Solving Eq.~\eqref{eq:critmanifold} for~$\tau$, we obtain~$\tau_{\rm sep.}(m^2,h^2)$. Choosing 
now~$\tau < \tau_{\rm sep.}$ for a given set of initial values~$(h^2_{\Lambda},m^2_{\Lambda})$, 
the theory necessarily approaches the regime
with broken chiral symmetry in the IR limit. For~$\tau > \tau_{\rm sep.}$, on the other hand, the theory remains in the
chirally symmetric regime. For a given value of the UV cutoff~$\Lambda$ and~$(h^2_{\Lambda},m^2_{\Lambda})$, the
quantity~$\tau_{\rm sep.}$ is therefore nothing but the dimensionless chiral phase transition temperature,~$  \tau_{\rm sep.}=\Tc/\Lambda$.
In fact,~$\Tc=\Lambda\tau_{\rm sep.}$ agrees with the result from Eq.~\eqref{eq:TcFourFPol}. Thus, our general statements in Sect.~\ref{sec:fermFP} 
concerning the interplay of the chiral and the deconfinement phase transition still hold.
\begin{figure*}[t]
\includegraphics[width=0.47\linewidth]{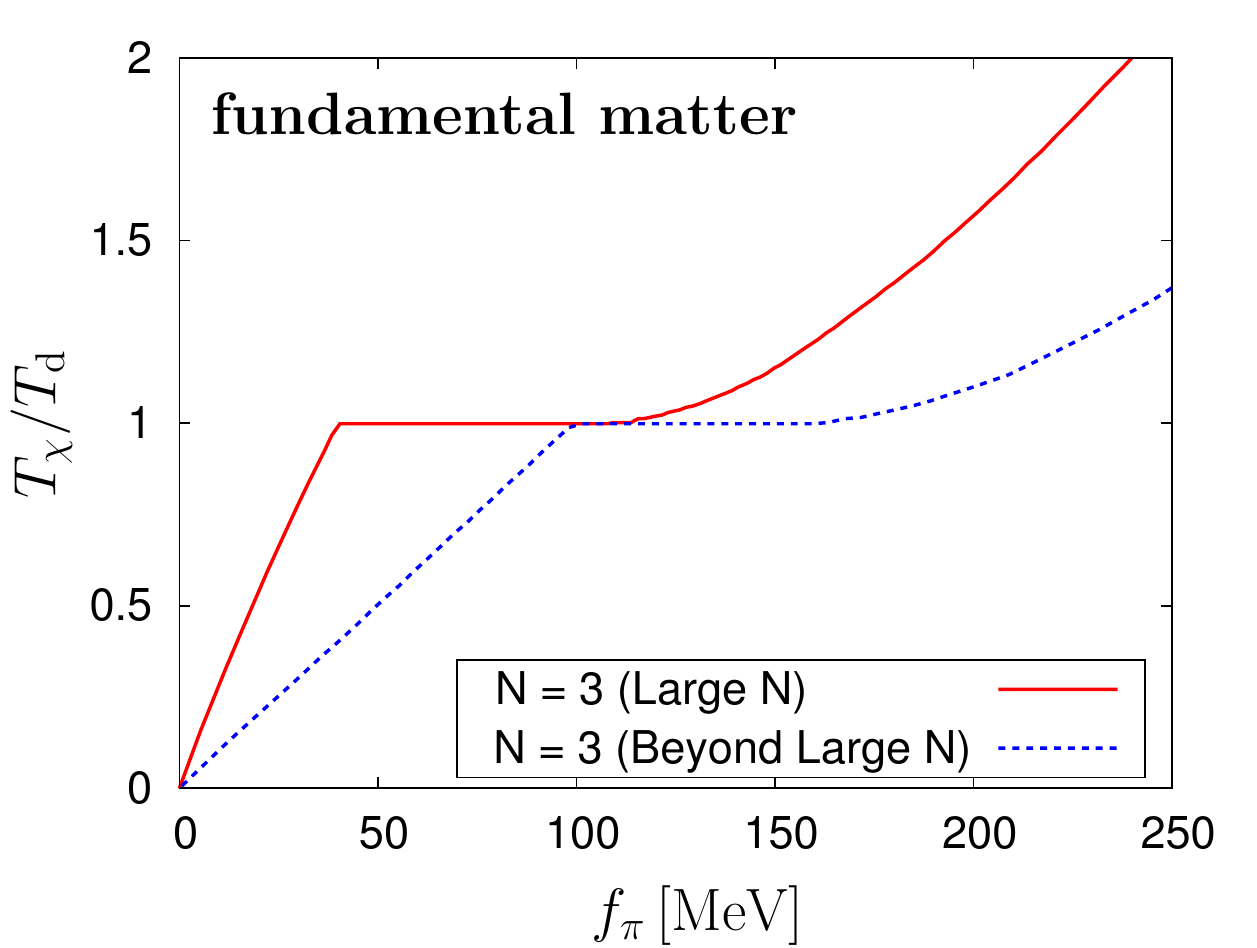}
\hspace*{0.8cm}
\includegraphics[width=0.47\linewidth]{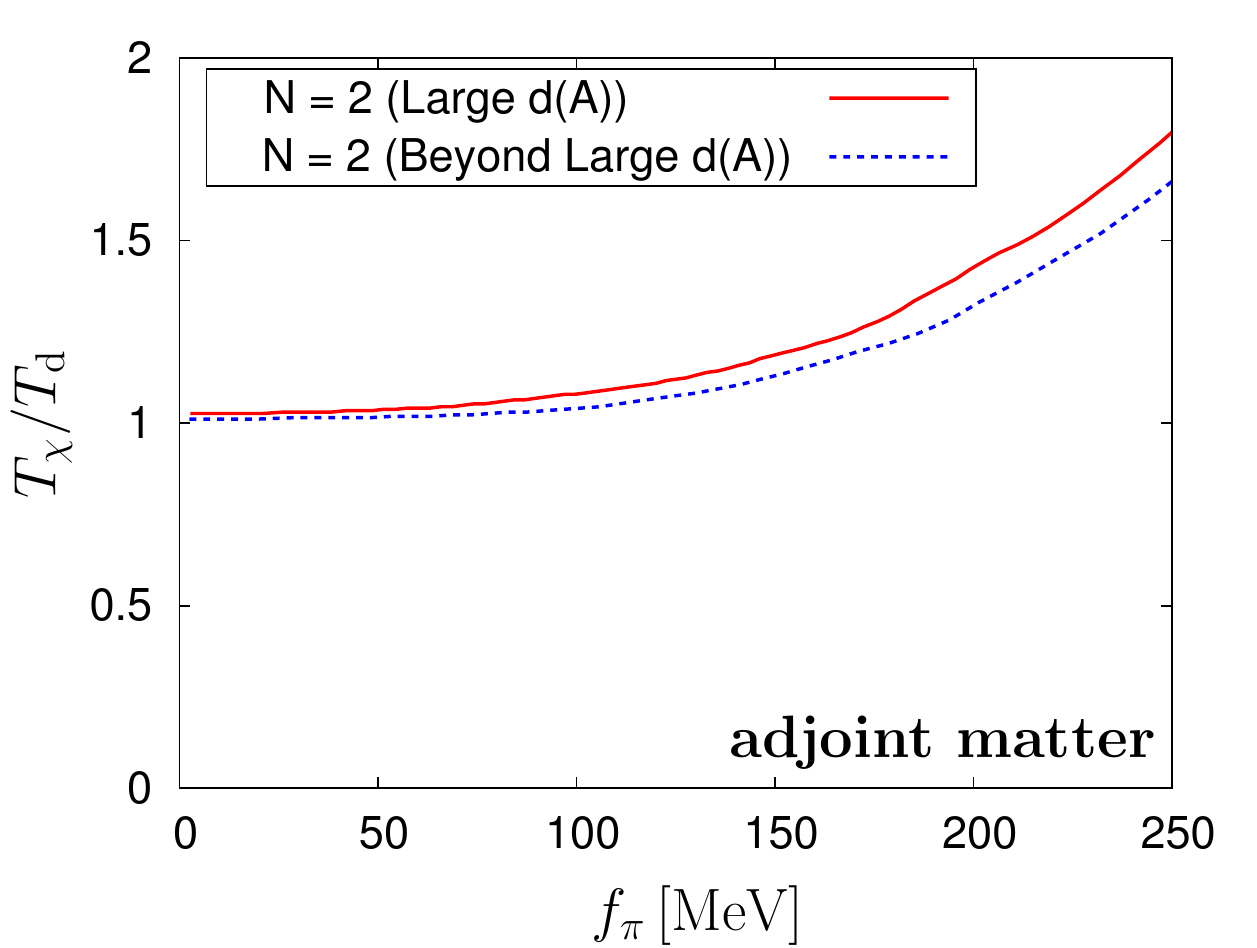}
\caption{ 
In the left panel, we show the phase diagram for two massless fundamental
quarks and $N=3$ in the plane spanned by the rescaled temperature $\Tc/\Td$ and
the value of the pion decay constant $f_{\pi}$
at $T=0$. 
In the right panel, the corresponding phase diagram for two massless quark flavors in the adjoint representation and~$N=2$
is shown. {In both panels, the results from the large-$d({\rm R})$ approximation are given
by the red (solid) line, whereas the blue (dashed) line depicts the results from our study including corrections 
beyond the large-$d({\rm R})$ limit.}
}
\label{fig:pdfundadj}
\end{figure*}

Let us now discuss how our phase diagrams in the~$(T,\luv)$-plane can be
translated into phase diagrams
in, e.~g., the~$(T,f_{\pi})$-plane. 
To this end, we need to follow the RG flow down to the long-range limit. As
discussed in Sect.~\ref{sec:setupmatter}, the mass parameter~$m^2$ assumes
negative values in the regime with broken chiral symmetry in the ground state
and the  vacuum expectation value~$\langle \Phi \rangle\equiv \bar{\Phi}_0$
becomes finite. It is therefore convenient to study the RG flow
of~$\bar{\Phi}_0$ and~$\bar{\lambda}_{\Phi}$ rather than that of~$\bar{m}^2$
and~$\bar{\lambda}_{\Phi}$. 
The flow equation of~$\bar{\Phi}_0$ can be obtained from the stationary condition:
\be
\frac{d}{dt}\left[\frac{\partial}{\partial\bar{\Phi}^2}\left(\frac{1}{2}\bar{m}^2\bar{\Phi}^2 + \frac{1}{8}\bar{\lambda}_{\Phi}\bar{\Phi}^4
\right)\right]_{\bar{\Phi}_0}\stackrel{!}{=}0\,.
\ee
To be specific, we find the
following RG flow equations for the regime with broken chiral symmetry in the ground state:
\be
\eta_{\Phi}&=& \frac{2}{3\pi^2}   \sum_{l=1}^{d({\rm R})}{\mathcal M}_{4,\perp}^{({\rm F})}(\tau,m_{\rm q}^2,\nu_l |\phi|) h^2 
\label{eq:etaLNBRO}\,,\\
\eta_{\psi}&=& 0\,,\\
\partial_t h^2 &=& (2\eta_{\psi} + \eta_{\Phi})h^2\,,\label{eq:h2LNBRO}\\
\partial_t \Phi_0^2 &=& -(\eta_{\Phi}\!+\! 2)\Phi_0^2 
\nn\\
&& \qquad\; 
- \frac{8}{\pi ^2}
\sum_{l=1}^{d({\rm R})} l_{1}^{\rm (F)}(\tau,m_{\rm q}^2,\nu_l |\phi|) \frac{h^2}{\lambda_{\Phi}},\label{eq:phi0flowLNBRO}
\\
\partial _t \lambda_{\Phi}&=& 2 \eta_{\Phi}\lambda_{\Phi}  -\frac{8}{\pi ^2}
\sum_{l=1}^{d({\rm R})} l_{2}^{\rm (F)}(\tau,m_{\rm q}^2,\nu_l |\phi|) h^4\,,\label{eq:lflowBRO}
\ee
where~$\Phi_0^2=k^{-2}Z_{\Phi}\bar{\Phi}_0^2$ and the (dimensionless) renormalized constituent quark mass
reads
\be
m_{\rm q}^2=h^2\Phi_0^2\,.\nn
\ee
In the following we will identify the pion decay constant~$f_{\pi}$ with~$Z_{\Phi}^{1/2}\bar{\Phi}_0$.
The (dimensionless) renormalized meson masses are given by
\be
m_{\pi}^2=0\quad\text{and}\quad m_{\sigma}^2=\lambda_{\Phi}\Phi_0^2\,.\nn
\ee
Since we are working in the large-$d({\rm R})$ limit in this section, the latter do not appear explicitly on the right
side of the flow equations. 

Recall that the scale for~$m_{\rm q}$ and~$m_{\sigma}$ is set by the symmetry breaking scale~$\ksb$ which is set by our
choice for~$h^2_{\Lambda}/m^2_{\Lambda}$. The role of the Yukawa coupling (as an additional parameter) becomes now
apparent from the relation
\be
m_{\sigma}^2 = \lambda_{\Phi}\Phi_0^2 \sim h^4 \Phi_0^2 \sim h^2 m_{\rm q}^2\,, \nn
\ee
which follows from
the flow equations of the couplings. Since the flow of the Yukawa coupling is not governed by the presence of 
a non-trivial IR attractive fixed-point, its value depends on~$\ksb$ and the initial value~$h_{\Lambda}$, as discussed above. 
Therefore the ratio~$m_{\sigma}^2/m_{\rm q}^2$ depends on our choice
for~$h_{\Lambda}$.
On the other hand, the initial value of the coupling~$\lambda_{\Phi}$ does not represent a free parameter 
of the theory. It is set to zero at~$k=\Lambda$ and therefore generated dynamically in the RG flow,
see also Eq.~\eqref{Eq:HSTAction}. 

Using the flow equations~\eqref{eq:etaphiLN}-\eqref{eq:lflow} and~\eqref{eq:etaLNBRO}-\eqref{eq:lflowBRO}, we can now
proceed and compute the phase diagram in the plane spanned by the temperature and the value of 
the pion decay constant at~$T=0$.
In Fig.~\ref{fig:pdfundadj} (left panel) we show our results for quarks in the fundamental representation and~$N=3$. 
For adjoint matter and~$N=2$, our results can be found in the right panel of Fig.~\ref{fig:pdfundadj}. 
To obtain these results, we have used~$\Lambda=1\,\text{GeV}$.
Moreover, we have again employed the data for the ground-state values of~$\bfe$ as 
obtained from a RG study of~SU($N$) Yang-Mills theories~\cite{Braun:2007bx,Braun:2010cy}.

In the case of fundamental matter and~$N=3$,
we observe that the upper end of the locking window ($\Td\approx \Tc$) roughly coincides with the physical value of the 
pion decay constant, provided that we fix the initial condition of the Yukawa coupling such that~$m_{\rm q}\approx 300\,\text{MeV}$
for~$f_{\pi}\approx 90\,\text{MeV}$, see left panel of Fig.~\ref{fig:pdfundadj}.
This observation is in accordance with results from lattice simulations and general expectations.
For~$f_{\pi} \lesssim {30}\,{\text{MeV}}$, we find~$\Tc < \Td$. More precisely, we observe that~$\Tc\sim f_{\pi}$ for small values of~$f_{\pi}$.
For~$f_{\pi}\gtrsim 100\,\text{MeV}$ 
($m_{\rm q}\gtrsim 350\,\text{MeV}$), we then have~$\Tc > \Td$. In this regime, the quarks are very heavy and the two phase transitions
are disentangled. 
Concerning the role of the Yukawa coupling, we find that the lower end of the locking window is shifted to smaller values of~$f_{\pi}$ when
we increase the initial value of the Yukawa coupling.
Moreover, we find that the size of the window does not strongly 
depend on our choice for~$h_{\Lambda}$. This is not unexpected
since we have found in our analysis of the fermionic fixed-point structure that the size of the locking window is solely
related to the value of the ratio~$h^2_{\Lambda}/m^2_{\Lambda}=\luv $. However, the translation of the upper and lower end of the locking window 
in $\luv$-space into values of physical observables does indeed depend on our choice
for both~$h_{\Lambda}$ as well as~$h_{\Lambda}^2/m_{\Lambda}^2$, as discussed above.

For adjoint matter and~$N=2$ as well as $N=3$, we find that~$\Tc > \Td$, even for very small values of~$f_{\pi}$.
We refer to Fig.~\ref{fig:pdfundadj} for our results for~$N=2$. 
{To obtain these results, we have used~$h_{\Lambda}=3$. However,~$\Tc > \Td$ holds for} 
arbitrary values of~$h_{\Lambda}$ in the large~$d({\rm R})$ limit, 
as suggested by our fermionic fixed-point analysis. In fact, our results in the large-$\dr$ limit are in 
accordance with our results in Fig.~\ref{fig:TlambdaF}, as it should be. 
For increasing~$f_{\pi}$, we observe that the chiral phase transition temperature increases further. Thus, 
we have~$\Tc > \Td$ for all values of~$f_{\pi}$. 

Finally we would like to add that it is also possible to tune the parameters~$h^2_{\Lambda}/m^2_{\Lambda}$
and~$m^2_{\Lambda}$ such that we obtain~$\Tc/\Td \approx 7.8$ {for~$N=3$}, 
as found in lattice simulations~\cite{Engels:2005te} of adjoint QCD without $\lambda_{\psi}$-deformation.
Of course, this requires that the UV cutoff~$\Lambda$ is adjusted 
to larger values in order to ensure that~$T/\Lambda$ is 
sufficiently small for the temperature range under consideration, see Ref.~\cite{Nishimura:2009me} 
for a PNJL model study in a mean-field approximation.

\subsection{RG Flow Beyond the Large-$\dr$ Approximation}\label{sec:BLNPBFP}
In the following we study the robustness of our results of the previous sections with respect
to~$1/\dr$-corrections. This includes an analysis of the role of Goldstone-mode 
fluctuations which are absent in the large-$d({\rm R})$ limit.

Our RG approach allows us to systematically include~$1/\dr$-corrections.
Due to the one-loop structure of the Wetterich equation, these corrections correspond to 1PI diagrams
with at least one internal boson line. In the chirally symmetric regime ($\Phi_0\equiv 0$), 
we then find the following set of equations:
\be
\eta_{\Phi}= \frac{2}{3\pi^2}   \sum_{l=1}^{d({\rm R})}{\mathcal M}_{4,\perp}^{({\rm F})}(\tau,0,\nu_l |\phi|) h^2 \,,
\label{eq:etaphiBLN} 
\ee
\be
\partial_t h^2 &=& (2\eta_{\psi} + \eta_{\Phi})h^2
 \nn\\
 && \; -\,\frac{2}{\pi ^2}\frac{1}{\dr}\sum_{l=1}^{\dr} 
 l_{1,1}^{\rm  (FB)}(\tau,0,\nu_l |\phi|,m^2)  h^4\,,\label{eq:h2BLN} \\
\partial_t m^2 &=& (\eta_{\Phi}\!-\! 2)m^2 - \frac{3}{2\pi^2} l_1(\tau,m^2) \lambda_{\Phi}\nn\\
&& \qquad +\frac{4}{\pi ^2}
\sum_{l=1}^{d({\rm R})} l_{1}^{\rm (F)}(\tau,0,\nu_l |\phi|) h^2,\label{eq:m2flowBLN} \\
\partial _t \lambda_{\Phi}&=& 2 \eta_{\Phi}\lambda_{\Phi} + \frac{3}{\pi^2} l_2(\tau,m^2) \lambda_{\Phi}^2 \nn\\
&& \qquad -\,\frac{8}{\pi ^2}\sum_{l=1}^{d({\rm R})} l_{2}^{\rm (F)}(\tau,0,\nu_l |\phi|) h^4\,,\label{eq:lflowBLN}
\ee
In regime with broken chiral symmetry~($\Phi_0 \neq 0$), the flow of the couplings is determined by the following 
equations:\footnote{In the flow equations for the Yukawa coupling and the bosonic wave-function renormalization,
we have dropped terms proportional to~$\Phi_0$. Concerning the Yukawa coupling,
it has been found that these terms only yield mild (quantitative)
corrections~\cite{Gies:2002hq,Gies:2009sv,Scherer:2012inprep}.
{With regard to} the bosonic wave-function renormalization, these
terms are of crucial importance for an accurate computation of
the critical exponents~\cite{Tetradis:1993ts,Berges:1997eu} which is beyond the
scope of the present work.}
\be
\eta_{\Phi}&=& \frac{2}{3\pi^2}   \sum_{l=1}^{d({\rm R})}{\mathcal M}_{4,\perp}^{({\rm F})}(\tau,m^2_{\rm q},\nu_l |\phi|) h^2 \,,
\label{eq:etaphiBLNbro} \\
\partial_t h^2 &=& (2\eta_{\psi} + \eta_{\Phi})h^2
 \nn\\
 && \; -\,\frac{1}{\pi ^2}\frac{1}{\dr}\sum_{l=1}^{\dr}\left[
 3\, l_{1,1}^{\rm  (FB)}(\tau,m^2_{\rm q},\nu_l |\phi|,m^2_{\pi}) \right. \nn\\
 &&
\left. \qquad\qquad\quad -  l_{1,1}^{\rm  (FB)}(\tau,m^2_{\rm q},\nu_l |\phi|,m^2_{\sigma})
 \right]h^4\,,\label{eq:h2BLNbro} \\
\partial_t \Phi_0^2 &=& -(\eta_{\Phi}+2)\Phi_0^2 + \frac{3}{2\pi^2}  l_1(\tau,m_{\sigma}^2)
+  \frac{3}{2\pi^2}  l_1(\tau,m_{\pi}^2) \nn\\
&& \qquad\qquad\quad -\frac{8}{\pi ^2}
\sum_{l=1}^{d({\rm R})} l_{1}^{\rm (F)}(\tau,m^2_{\rm q},\nu_l |\phi|) \frac{h^2}{\lambda_{\Phi}},\label{eq:m2flowBLNbro}
\\
\partial _t \lambda_{\Phi}&=& 2 \eta_{\Phi}\lambda_{\Phi} + \frac{9}{4\pi^2} l_2(\tau,m_{\sigma}^2) \lambda_{\Phi}^2
 + \frac{3}{4\pi^2} l_2(\tau,m_{\pi}^2) \lambda_{\Phi}^2 \nn\\
&& \qquad\qquad\quad -\,\frac{8}{\pi ^2}\sum_{l=1}^{d({\rm R})} l_{2}^{\rm
(F)}(\tau,m^2_{\rm q},\nu_l |\phi|) h^4\,.\label{eq:lflowBLNbro}
\ee
The threshold functions\footnote{Note that the functions~$l_1^{\rm (B)}$, $l_2^{\rm (B)}$
and~$l_{1,1}^{\rm (FB)}$ depend implicitly on~$\eta_{\Phi}$.} 
can be found in App.~\ref{app:thresfcts} and Ref.~\cite{Braun:2008pi}.
For simplicity, we do not include the running of the fermionic wave-function renormalization in the present study, although
it can be taken into account straightforwardly, as illustrated in, e.~g., Refs.~\cite{Berges:1997eu,Gies:2002hq,Braun:2008pi,Braun:2009si}
for the case~$\bfe=0$. {As discussed in the previous subsection, this} is {\it not} an approximation in the large-$\dr$ limit.
Beyond the large-$\dr$ limit, it has been found in Refs.~\cite{Berges:1997eu,Gies:2002hq,Braun:2008pi,Braun:2009si}
that the anomalous dimension~$\eta_{\psi}$ is still small. This can be traced back to the fact that the running of~$Z_{\Phi}$ is solely governed
by 1PI diagrams with at least one internal boson and fermion line. Such diagrams are parametrically suppressed 
in the regime with broken chiral symmetry due to the large mass of the fermions, but they are also suppressed in the chirally symmetric 
regime due to the large mass of the bosons.
As a consequence, the running of~$Z_{\psi}$ only yields mild corrections to the symmetry breaking scale~$\ksb$. 
In the following we will only take into account $1/\dr$-corrections in those RG equations 
which are also non-zero in the large-$\dr$ limit. The inclusion of the running of~$Z_{\psi}$ is left to future work.

{Using the flow} equations of the Yukawa coupling and the bosonic mass parameter in the chirally symmetric regime, we
can study again the RG flow of the ratio~$h^2/m^2$. We now find 
\be
&&\partial_t \left(\frac{h^2}{m^2}\right)=(2\!+\! 2\eta_{\psi})\left(\frac{h^2}{m^2}\right)
\!+\! \frac{3}{2\pi^2} l_1(\tau,m^2) \lambda_{\Phi}  \left(\frac{h^2}{m^4}\right) 
 \nn\\ 
 &&\qquad\qquad\qquad\quad
 - \frac{4}{\pi ^2}
\sum_{l=1}^{\dr} l_{1}^{\rm (F)}(\tau,m^2_{\rm q},\nu_l |\phi|)\left(\frac{h^2}{m^2}\right)^2 \nn\\
&& \qquad\;
-\frac{2}{\pi ^2}\frac{1}{\dr}\sum_{l=1}^{\dr} 
 l_{1,1}^{\rm  (FB)}(\tau,m^2_{\rm q},\nu_l |\phi|,m^2)  \left(\frac{h^4}{m^2}\right).
\label{eq:lambdapsiBLN}
\ee
Using~$\lambda_{\psi}\equiv (h^2/m^2)$, the first term on the right-hand side 
as well as the terms in the second line can be straightforwardly identified with terms appearing in
the RG equation of the four-fermion coupling~$\lambda_{\psi}$, see Eq.~\eqref{eq:lpsi_flow}. 
These are the leading order terms of the large-$\dr$ expansion.
The second term 
on the right-hand side corresponds to a $1/\dr$-correction {and effectively couples} the 
flow of~$h^2/m^2$ ($\sim$ four-fermion coupling) to the flow of the four-boson coupling ($\sim$ 8-fermion coupling). 
Since it can be shown that the RG flow of fermionic self-interactions is fully decoupled in the 
point-like limit~\cite{Braun:2011pp}, this term resolves (part of) the momentum dependence of
the four-fermion interaction.
The expression in the third line also represents a $1/\dr$-correction and can be traced back to the
running of the Yukawa coupling. Without the terms in the third line, it is not possible
to reproduce the prefactor of the term~$\sim \lambda_{\psi}^2$ in the flow equation~\eqref{eq:lpsi_flow}
in the limit $m^2\gg 1$ (point-like limit). As pointed out in Ref.~\cite{Braun:2011fw},
this can be seen immediately from the following relation
\be
l_{1,1}^{\rm  (FB)}(\tau,m^2_{\rm q},\nu_l |\phi|,m^2) 
\stackrel{ (m\gg 1)}{\longrightarrow}\frac{1}{m^2}\, l_{1}^{\rm (F)}(\tau,m^2_{\rm q},\nu_l |\phi|)\,.\nn
\ee
For finite~$m^2$, the expression in the third line on the right-hand side of Eq.~\eqref{eq:lambdapsiBLN} also resolves
part of the momentum structure of the four-fermion vertex beyond the point-like limit.

Let us now discuss our results for the phase diagrams in the $(T,f_{\pi})$-plane beyond the large-$\dr$ limit.
In Fig.~\ref{fig:pdfundadj}, we show our results for quarks in the fundamental representation and~$N=3$ as well as
for quarks in the adjoint representation and~$N=2$. We have chosen these representations and values 
for~$N$ since they play a prominent role from a phenomenological point of view. 
For quarks in the fundamental representation, we observe
that our results agree with those from our large-$N$ study, at least on a
qualitative level.\footnote{Note that we have
fixed the initial value of the Yukawa coupling by requiring that~$m_{\rm q}\approx 300\,\text{MeV}$ for
$f_{\pi}\approx 90\,\text{MeV}$. The same initial value for the Yukawa coupling has then been used to compute
the phase transition temperature for all other values of~$f_{\pi}$ as well. Thus, we have only varied
the initial value of the bosonic mass parameter~$m^2$ to change the value of~$f_{\pi}$. Recall 
that~$\lambda_{\psi}\sim h^2/m^2$.} This means we still
have three distinct regimes: one regime with~$\Tc <\Td$ for small values of~$f_{\pi}$, one regime 
with~$\Tc\approx \Td$ (locking window), and a regime with~$\Tc > \Td$ for large values of $f_{\pi}$. Also, the
size of the locking window is roughly the same as in the {large-$N$ approximation.} However, the lower and
the upper end of the window have been shifted to larger values of~$f_{\pi}$. The locking window begins
at~$f_{\pi}\approx 100\,\text{MeV}$ and ends at~$f_{\pi}\approx 150\,\text{MeV}$. Thus, the physical value 
of the pion decay constant is slightly below the lower end of the locking window. 

For quarks in the adjoint representation and~$N=2$, we find that our results are less strongly 
affected by corrections arising beyond the {large-$d({\rm A})$ approximation, see}
Fig.~\ref{fig:pdfundadj} (right panel).
To be specific, we observe that~$\Tc > \Td$ for~$f_{\pi} >0$, even if we take $1/\dr$-corrections into account. 
The results only differ with respect to the slope of the chiral phase transition temperature as a function
of~$f_{\pi}$. As in the case of fermions in the fundamental representation, the slope is steeper in the
{large-$d({\rm A})$ limit.} We conclude that fluctuations of the Nambu-Goldstone modes 
tend to lower the sensitivity of~$\Tc$ on~$f_{\pi}$.

The results for adjoint quarks in Fig.~\ref{fig:pdfundadj} have been obtained by choosing~$h_{\Lambda}=3$ for the initial
value of the Yukawa coupling. The value of the pion decay constant can then be
varied by varying only the initial value of the bosonic mass
parameter~$m_{\Lambda}$. As in the case of fundamental quarks, it is in
principle possible to fix the initial condition for the Yukawa coupling
by requiring that the constituent quark mass assumes a given value for a given value of the pion decay constant.
For adjoint quarks, we refrain from fixing the initial condition~$h_{\Lambda}$ in this way but rather illustrate how our
results depend on the choice for~$h_{\Lambda}$, see Fig.~\ref{fig:pdadjhdep}. We observe that the dependence of~$\Tc$
on~$f_{\pi}$ becomes stronger for larger values of~$h_{\Lambda}$. Most importantly, however, we find 
that~$\Tc>\Td$ for~$N=2$, independent
of our choice for~$h_{\Lambda}>0$. We stress that the mechanism underlying this observation is 
the deformation of the (fermionic) fixed-point structure due to the presence of the confining gauge dynamics.
\begin{figure}[t]
\includegraphics[width=1\linewidth]{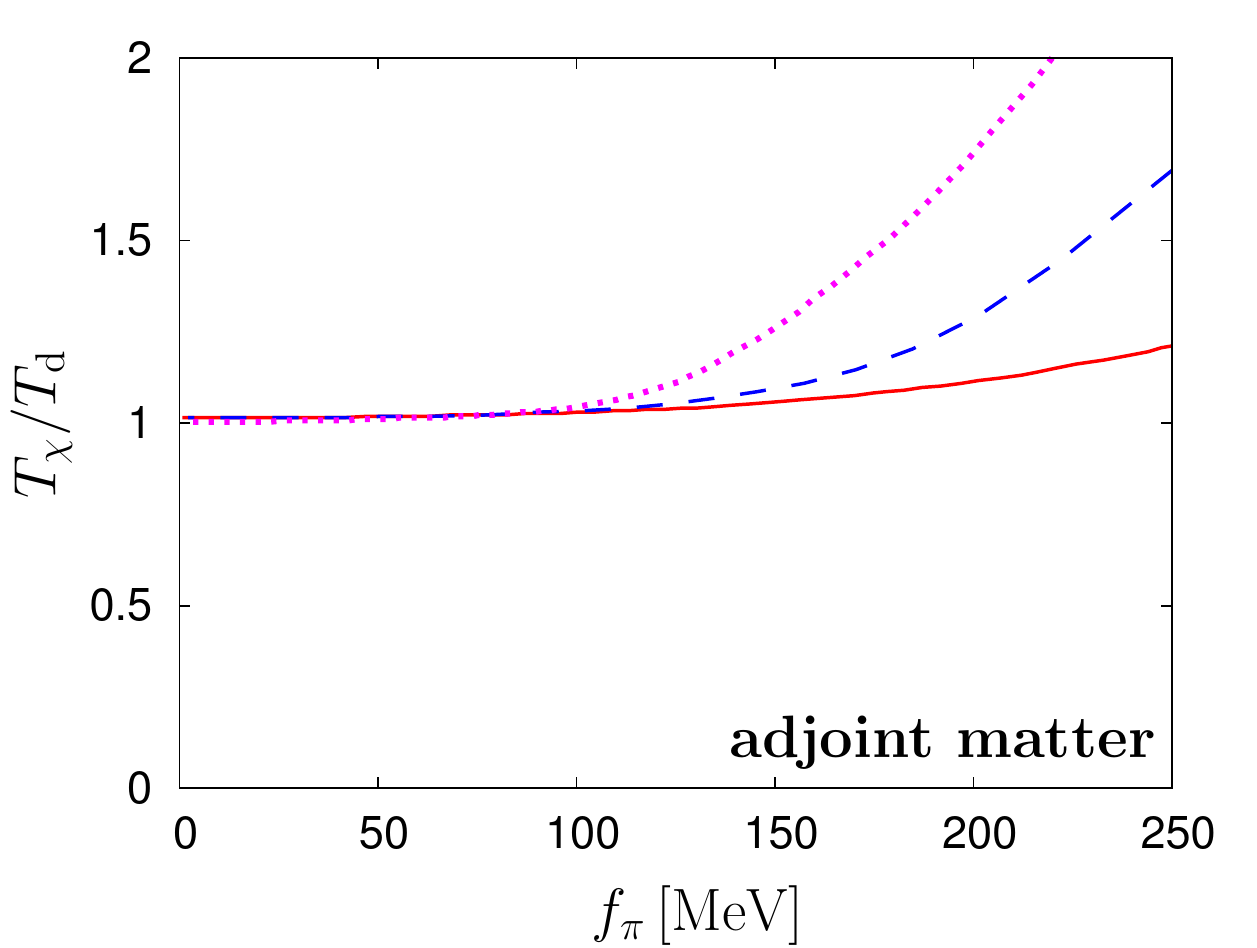}
\caption{
Ratio $\Tc/\Td$ of the chiral and the deconfinement phase transition temperature as a function
of the zero-temperature value of the pion decay constant~$f_{\pi}$ for two massless adjoint quarks 
and~$N=2$. The various lines illustrate the dependence of our
results on the initial condition (UV value) for the Yukawa coupling. The results have been obtained
for~$h_{\Lambda}=2,3,4$ (from bottom to top).
}
\label{fig:pdadjhdep}
\end{figure}

Let us finally comment on the order of the chiral phase transition in the $(T,f_{\pi})$ phase diagram. In 
Ref.~\cite{Braun:2011fw}, it was found for fundamental fermions and~$N=3$
that the chiral phase transition is of first order within the locking 
window. To be more precise, we observe that the chiral phase transition is of first order 
for $100\,\text{MeV}\lesssim f_{\pi} \lesssim 150\,\text{MeV}$ for $N=3$. Above and below the locking 
window, the chiral phase transition is of second order. In particular, the observation of a first-order
region might be a shortcoming of our approximations: we have simply used the data for~$\bfe(T)$ from
a study of pure SU($3$) Yang-Mills theory, but neglected the back-reaction of the matter sector
on the confinement order parameter. Within the locking window, the first-order phase transition in the
gauge sector induces a first-order chiral phase transition. As argued in Ref.~\cite{Braun:2011fw},
a first-order chiral transition may still occur in the $(T,f_{\pi})$-plane, even if we go beyond the present 
approximation. However, this would then require that the confinement order parameter rises rapidly 
for~$T\gtrsim \Td$. A test of this conjecture is beyond the scope of the present work and left to future studies.

For adjoint matter and~$N=2$, we observe that~$\Tc > \Td$ for all values of~$f_{\pi}>0$. Therefore
the dynamics at the chiral phase transition is less affected by the confining dynamics. 
Loosely speaking, the latter only pushes~$\Tc$ above~$\Td$. 
Within the present approximation, we therefore find that the chiral phase transition is of second
order for all values of~$f_{\pi}>0$. This result is consistent with lattice simulations for~$N=2$, see 
Refs.~\cite{Karsch:1998qj,Engels:2005te}.

\section{Conclusions and Outlook}\label{sec:conc}
In the present paper we have analyzed the interplay of the chiral and the deconfinement phase transition in
gauge theories with matter fields in different representations, with an emphasis on quarks in the 
fundamental and the adjoint representation. To this end, we have computed phase diagrams in the
plane spanned by the temperature and the pion decay constant using a simple ansatz for the quantum
effective action. This ansatz allowed us to study the fixed-point structure in the matter sector analytically. In particular,
it opened up the possibility to analyze the impact of the confinement order parameter on
the chiral fixed-point structure. The latter is directly related to the order parameter for chiral symmetry breaking. 

For theories with quark fields living in a given representation~R,
we have found that the interplay of the chiral and the deconfinement
phase transition clearly depends on the sign of the quantity~$\tr_{\rm R}L_{\rm R}[\bfe]$ in the center symmetric 
phase. The relation of~$\tr _{\rm R}L_{\rm R}[\bfe]$ to the standard Polyakov loop (for a given representation R) has been discussed
in detail in Sect.~\ref{sec:GA}. To be specific, our fixed-point analysis suggests that~$\Tc > \Td$ for adjoint quarks, at least in the large-$\dr$ limit.
This observation is in accordance with results from lattice simulations~\cite{Karsch:1998qj,Engels:2005te,Bilgici:2009jy}.
For quarks in the fundamental representation, our findings are also compatible with 
lattice QCD studies~\cite{Cheng:2006qk,Aoki:2006br,Aoki:2006we,Aoki:2009sc},
first-principles continuum
studies~\cite{Braun:2009gm,Pawlowski:2010ht,Fischer:2011mz}, and earlier
analytic (model) studies~\cite{Coleman:1980mx,Meisinger:1995ih,Braun:2011fw}.

We have also investigated how robust our predictions for the~$(T,f_{\pi})$ phase diagram
are, once $1/\dr$-corrections are taken into account in the matter sector.
Such corrections are associated with fluctuations of the Nambu-Goldstone modes of the theory. 
For quarks in the fundamental representation, we have found that
the locking window ($\Tc\approx\Td$) is shifted to larger values of~$f_{\pi}$ but remains finite. At the physical 
point~$(f_{\pi}\approx 90\,\text{MeV})$, we have~$\Tc\lesssim \Td$. For adjoint quarks and~\mbox{$N=2$}, we have found
that~$\Tc > \Td$ for all values of~$f_{\pi}>0$, even if~$1/\dr$-corrections are
taken into account. 
In this respect, the finite-temperature dynamics of gauge theories with adjoint matter appear to be
distinct from gauge theories with fundamental matter, at least for~$N=2$.
Recall that this observation is also consistent with lattice studies 
of SU($2$) gauge theory with two flavors of adjoint quarks~\cite{Karsch:1998qj,Engels:2005te,Bilgici:2009jy}.

We would like to add that our fixed-point analysis can also help to guide the
development of QCD low-energy models in the future. 
To be specific, we have used the order-parameter potential spanned by
the background temporal gauge field. Moreover, we have not employed the
assumption~$\tr_{\rm R}L_{\rm R}[\bfe]=\langle \tr_{\rm R} L_{\rm
R}[A_0]\rangle$ which is often used in PNJL/PQM-type model studies.
Instead, we have considered the quantity~$\tr_{\rm R} L_{\rm R}[\bfe]$ in the
present study.
This corresponds to working in a specific class of gauges, namely the class of 
Polyakov-DeWitt gauges. Our study
suggests that the assumption~$\tr_{\rm R}L_{\rm R}[\bfe]=\langle \tr_{\rm R}L_{\rm R}[A_0]\rangle$ 
is justified in a mean-field approximation $(N\to\infty)$.
For finite~$N$ (and, in particular, for quarks in representations other than the fundamental one), however,
{the situation may change.} Depending on the representation~R, the sign
of~$\tr_{\rm R}L_{\rm R}[\bfe]$ 
can be different below and above the phase transition,
whereas~$\langle \tr_{\rm R}L_{\rm R}[A_0]\rangle$ can be defined to be positive for all temperatures.

Of course, the present analysis can be improved in many ways. For example, one may consider to take into account
the back-reaction of the matter fields on the quantity~$\tr_{\rm R}L_{\rm R}[\bfe]$. Such contributions will 
push~$\tr_{\rm R}L_{\rm R}[\bfe]$ to larger values in the low-temperature phase. We expect that this will weaken the 
mechanisms governing the dynamics in our
present study.  For fundamental quarks, for example, this may shrink the locking window. For adjoint quarks, on the other hand,
the quantity~$\tr_{\rm R}L_{\rm R}[\bfe]$ may still be negative over a wide range of temperatures. Therefore, $\Tc >\Td$
may persist for~$N=2$, even if we take these back-reactions into account. In any case, we have presented
{first} predictions for the~$(T,f_{\pi})$ phase diagram and our
analysis reveals a simple mechanism governing the interplay
of the chiral and the deconfinement phase transition. It would be interesting to see whether and how this
mechanism persists in the presence of a finite quark chemical potential and/or a finite external magnetic 
field. The latter deforms the fermionic fixed-point structure in a way~\cite{Scherer:2012nn,Fukushima:2012xw}  
which is indeed reminiscent of the deformation discussed here for adjoint quarks.

\emph{Acknowledgments.}
The authors are very grateful to H.~Gies,  J.~M.~Pawlowski and B.-J.~Schaefer 
for useful discussions and critical comments on the manuscript. Moreover, the authors
thank R. Alkofer, A.~Janot, D.~D.~Scherer and A.~Wipf for useful discussions.
{JB acknowledges support by the DFG research training group GRK 1523/1,}
and by the Helmholtz International Center for FAIR within the LOEWE program of the State of Hesse.
TKH is recipient of a DOC-fFORTE-fellowship of the Austrian Academy of Sciences and supported by the
FWF doctoral program DK-W1203-N16.

\appendix
\section{Threshold functions}\label{app:thresfcts}
In the computation of the RG flow equations, a regulator function needs to be specified
which determines the regularization scheme~\cite{Wetterich:1992yh}.
Here, we have used a linear spatial regulator function for the bosonic as well 
as for the fermionic degrees of freedom~\cite{Litim:2000ci,Litim:2001fd,Litim:2001up,Litim:2006ag,Blaizot:2006rj}. 
To be specific, we have chosen
\be
\!\!\!\!\!\!\! R_{\rm B}(\vec{p}^{\,2})=\vec{p}^{\,2}\left(\frac{k^2}{\vec{p}^{\,2}}\!-\!1\right)\theta(k^2\!-\!\vec{p}^{\,2})
\equiv \vec{p}^{\,2} r_{\rm B} \left({ \frac{\vec{p}^{\,2}}{k^2}}\right),
\label{eq:bosreg}
\ee
for the bosons, whereas we have chosen
\be
\!\!\!\!\!\!\! R_{\psi}(\vec{p})=\vec{p}\fslash\left(\sqrt{\frac{k^2}{\vec{p}^{\,2}}}\!-\!1\right)\theta(k^2\!-\!\vec{p}^{\,2})
\equiv \vec{p}\fslash\, r_{\psi} \left({ \frac{\vec{p}^{\,2}}{k^2}}\right)
\label{eq:fermreg}
\ee
for the fermionic degrees of freedom. 

Now we define the threshold function~${\mathcal M}_{4,{\perp}} ^{({\rm F})}$. This function 
represents a 1PI diagram with two internal fermion lines and contributes 
to the RG flow of the bosonic wave-function renormalization. 
Those threshold functions, which are not defined in this appendix, 
can be found in Refs.~\cite{Braun:2008pi,Braun:2011fw,Braun:2011pp}. 

To define the threshold function~${\mathcal M}_{4,{\perp}} ^{({\rm F})}$, it is convenient to introduce a 
dimensionless propagator for the fermions:
\be
\tilde{G} _{\psi} (x_0,\omega)=\frac{1}{ x_0 + x(1+r_{\psi})^2 + \omega}\,,
\ee
where $x=\vec{p}^{\,2}/k^2$. In terms of this propagator, the threshold function
entering the anomalous dimension of the bosons can be written as follows:
\be
&&{\mathcal M}_{4,{\perp}} ^{({\rm F})}(\tau,\omega,\mu)\nn\\
&&=(d\!-\!1)\tau\sum_{n=-\infty}^{\infty} \int _0 ^{\infty} \!dx x^{\frac{d-3}{2}}\tilde{\partial}_t
\Bigg\{ x (1\!+\! r_{\psi}) \tilde{G} _{\psi} (x_0^{\psi},\omega)\!\times\nn\\
&&\qquad\times\Bigg[\frac{2x}{d-1} \left(\frac{d^2}{dx^2}(1+r_{\psi}) \tilde{G} _{\psi} (x_0^{\psi},\omega)\right)\nn\\
&&\qquad\qquad + \frac{d+1}{d-1} \left(\frac{d}{dx}(1+r_{\psi}) \tilde{G} _{\psi} (x_0^{\psi},\omega)\right)\Bigg]\nn\\
&& \quad + x_0^{\psi}  \tilde{G} _{\psi} (x_0^{\psi},\omega) 
\Bigg[
\frac{2x}{d-1} \left( \frac{d^2}{dx^2} \tilde{G} _{\psi} (x_0^{\psi},\omega) \right)\nn\\
&& \quad\qquad\qquad\qquad\qquad\qquad\; +  \left( \frac{d}{dx} \tilde{G} _{\psi} (x_0^{\psi},\omega) \right)
\Bigg]\Bigg\},
\ee
where~$x_0^{\psi}=(\tilde{\nu}_n+ 2\pi\tau\mu)^2$ and~$\tilde{\nu}_n=(2n+1)\pi\tau$.
Here, the derivative~$\tilde{\partial}_t$ with respect to the regulator function is defined as follows:
\be
\tilde{\partial}_t 
&=& \frac{1}{x^{1/2}}\theta(1-x)\frac{\partial}{\partial r_{\psi}}\,.
\ee
We do not display terms~$\propto \eta_{\psi}$ since we have not taken into account these
contributions in our numerical analysis.

\bibliographystyle{h-physrev3}
\bibliography{bib_source}

\end{document}